\documentclass[12pt,preprint]{aastex}

\begin{document}

\title{A photoionized Herbig-Haro object in the Orion nebula}

\author{K. P. M. Blagrave and P. G. Martin}
\affil{Department of 
Astronomy and Astrophysics and Canadian 
Institute for Theoretical Astrophysics, University of Toronto,
60 St. George Street, Toronto, ON M5S 3H8, Canada}
\email{blagrave@cita.utoronto.ca, pgmartin@cita.utoronto.ca}

\and

\author{J. A. Baldwin}
\affil{Physics and Astronomy Department, 3270 Biomedical Physical Sciences Building, \\
Michigan State 
University, East Lansing, MI 48824}
\email{baldwin@pa.msu.edu}

\begin{abstract}
The spectra of Herbig Haro objects are usually characteristic of
ionization and excitation in shock-heated gas, whether an internal shock
in an unsteady outflow or a bow shock interface with the interstellar   
medium.
We examine the eastern-most shock -- the leading optically visible shock -- of a
Herbig Haro outflow (HH~529) seen projected on the face of the Orion Nebula, using deep
optical echelle spectroscopy, showing
that the spectrum of this gas is consistent with photoionization by
$\theta^1$ Ori C.
By modeling the emission lines, 
we determine a gas-phase abundance of Fe which is consistent with the
depleted (relative to solar) abundance found in the Orion nebula --
evidence for the presence of dust in the nebula and therefore in
the Herbig Haro outflow.
The spectrum also allows for the calculation of 
temperature fluctuations, $t^2$, in the nebula and the shock.  These fluctuations 
have been used to explain discrepancies
between abundances obtained from recombination lines versus those obtained from
collisionally-excited lines, although to date there has not been a robust theory for
how such large fluctuations ($t^2 > 0.02$) can exist.
\end{abstract}


\section{Introduction \label{intro}}
Star-forming regions -- such as the Orion Nebula -- are home to
various phenomena associated with the early stages of stellar evolution.
Some of the more prominent features in the visible part of the spectrum 
are
the arcs associated with gas flows known as Herbig Haro (HH) flows
\citep{rei01}.
Many of these flows have been identified in the Orion Nebula and have had
both their radial \citep{doi04} and tangential \citep{doi02} velocities
measured.  The origins of these flows have in a few cases been associated
with IR sources embedded within the Orion Molecular Cloud 1 South (OMC-1S)
\citep{doi02}.
However,
there are many flows that have not been paired with any source (X-ray,
radio, or near-IR) -- including HH~529 \citep{ode03}.
This flow
contains at least three curved shocks which appear in [O~{\sc iii}] WFPC2 
images
\citep{ode96} and extend approximately $36\arcsec$
from the centre of the inferred source of the optical outflow (OOS) at
$\alpha$, $\delta$ (J2000) = $5^{\rm h}35^{\rm m}14\fs56$,$-5^{\rm
o}$23\arcmin54\arcsec \citep{ode03,doi04}.\footnote{
\footnotesize{
In addition to HH~529, many other HH flows (HH~269, HH~202
and HH~203/204) appear to originate in the OOS region -- supplying ample
evidence for OOS housing HH flow driver(s).
\citet{smi04} have detected an infrared source (IR source 2 in their 
Table~2; $\alpha$, $\delta$
= $5^{\rm h}35^{\rm m}14\fs40$,$-5^{\rm o}$23\arcmin51\farcs0) which lies
within $3\arcsec$ of the predicted location of the OOS.  \citet{zap04} 
have also observed this
source at 1.3cm.
Within the OOS region,
\citet{bal00b} have identified another
near-IR source (`s' in their Fig.~20) which was concurrently labelled
HC209 \citep{hil00}: $\alpha$, $\delta$ = $5^{\rm h}35^{\rm
m}14\fs57$,$-5^{\rm o}$23\arcmin50\farcs8.
Recently, an X-ray source (F421,
\citet{fei02}) has been found to be coincident with this near IR source.
However, there is still not definitive proof as to the particular
driving source as neither of these sources lies directly in line with the 
flow of HH~529.
}
}
This is 0.08~pc in the plane of sky given a distance to the nebula of 460~pc 
(\citet{bal00b}, hereafter BOM).

The radial (line-of-sight) velocity is $-44$~km~s$^{-1}$
\citep{doi04}.
This radial velocity is quoted as ``systemic''
-- relative to the [O~{\sc iii}] nebular component, which itself has a
heliocentric
velocity of $+18\pm2$~km~s$^{-1}$ \citep{doi04}.  Coupling this with the
heliocentric radial velocity of the PDR ($+28$~km~s$^{-1}$,
\citet{gou82}), we obtain
a radial velocity relative to the source embedded within OMC-1: $-54$~km~s$^{-1}$.
The 
average proper motion velocity is $54$~km~s$^{-1}$ \citep{doi04} which leads 
to a total velocity of $76$~km~s$^{-1}$ (with respect to OMC-1S) at an angle of
$45^{\rm o}$ out of the plane of the sky.

Using this geometry, a distance from the embedded source to the leading edge of 
the
eastern-most shock can be calculated: 0.12~pc ($36\arcsec\times\sqrt{2}$).
Assuming that the source lies within OMC-1S and that $\theta^1$~Ori~C
is itself $\sim0.25$~pc from the main ionization front \citep{wen95,ode01b}, this 
would place
the HH~529 system on
the far side (i.e., further from the observer)
of $\theta^1$~Ori~C.
It is remarkable that the flow
has emerged from the cloud into the ionized zone.
 
The dynamical age of the HH~529 system can be calculated from the average 
proper motion
(25 mas yr$^{-1}$).
Assuming that the proper motion has remained
constant over the $36\arcsec$ from its point of origin, we
find the dynamical age of the eastern-most visible feature of HH~529 to be 
roughly 1500
years.  All shock model timescales will need to be consistent with this dynamical age
in order for the model to be valid (\S~\ref{anal.model}).

As was recognized by \citet{ode97a}, the fact that this and other Orion 
nebula
HH flows
appear strongly in [O~{\sc iii}] (atypical of most HH flows which show 
much
lower ionization)
suggests that these shocks are photoionized.  We examine the physical
conditions of HH~529 by comparing our high-resolution echelle spectra with
(matter-bounded) photoionization models of this feature.
Other studies of non-photoionized HH flows show evidence for
a decrease in the amount of Fe depletion in some of the shocks, as 
determined
from [Fe~{\sc ii}] lines \citep{boh01,bec96}.
This has been linked to grain
destruction as matter originating from the molecular cloud passes
through the shocks.
In this paper, we assess the Fe depletion using
a set of [Fe~{\sc iii}]
lines in the eastern-most feature of HH~529.

\section{Observations \label{observe}}

\begin{figure}
\epsscale{1.0}
\centerline{\scalebox{1.0}[1.0]{\plotone{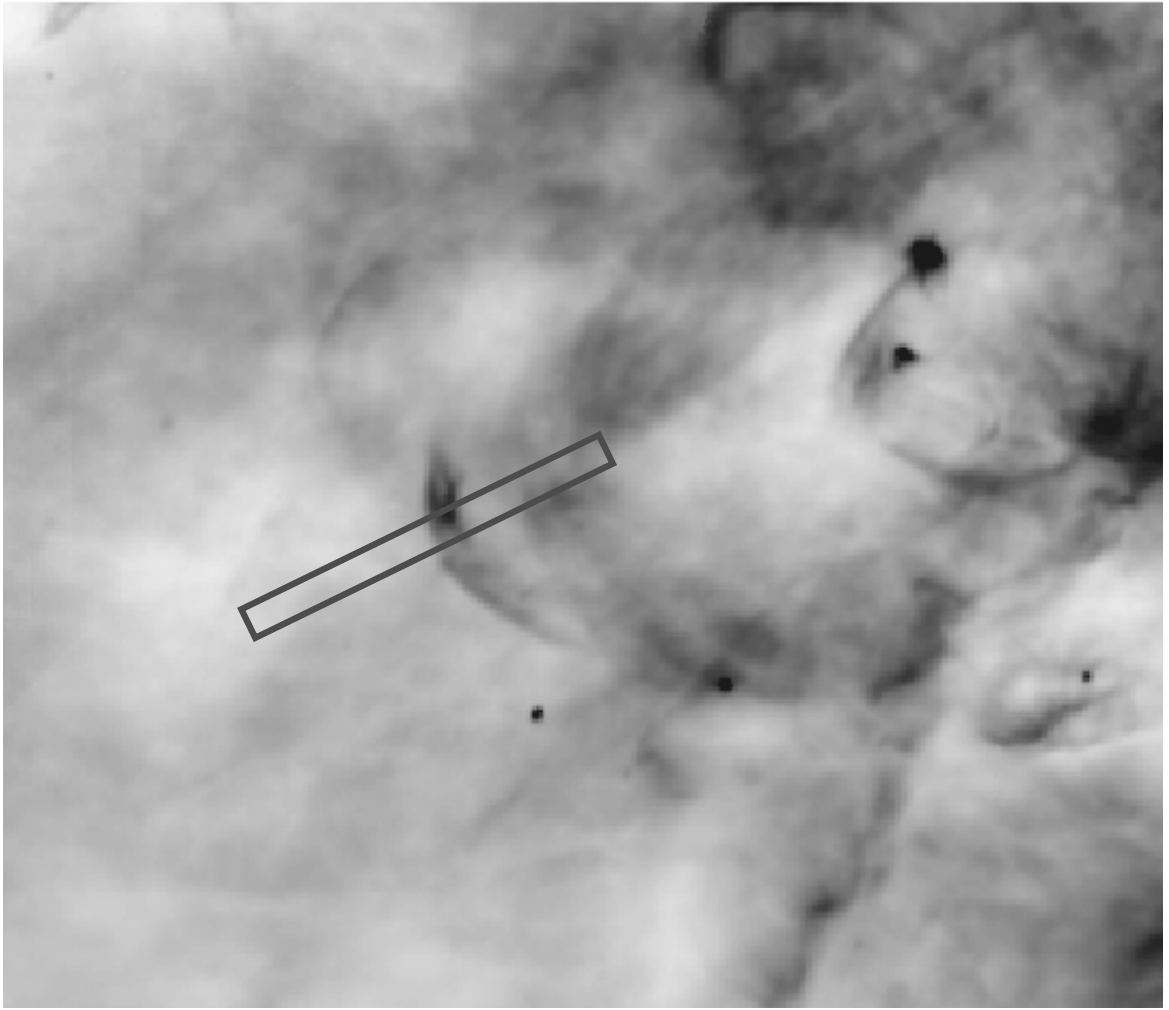}}}
\caption{Red x2 slit position as determined from a Polaroid of the slit
taken during observations, and the surface brightness variation across the slit
as compared with the underlying F656N (H$\alpha$)
WFPC2 image \citep{ode96}. The slit is $12\farcs5 \times 1\arcsec$.
\label{slitpos}}
\end{figure}

Spectra were obtained using the echelle spectrograph on the 4m Blanco 
telescope at CTIO (see \citet{bal00} for details) covering the spectral 
range from the near-UV ($3500$\AA) to the near-IR ($7500$\AA).  Three 
sets 
of red 
and blue spectra were 
obtained on two dates in 1997 and 1998.  One of the three slit positions 
(x2, see Fig.~\ref{slitpos}) intentionally
overlaps with the eastern-most visible feature of HH~529.
Wavelength and flux calibrations were performed as in \citet{bal00}.
We also used archival flux-calibrated (\citet{ode99} using \citet{bal91}) HST WFPC2 
associations
(F487N, F502N, F547N, F631N, F656N, F658N, F673N) and
Bally mosaics of these associations (less F487N).
As there were a series of discrepant exposure times in the image headers of the
Bally mosaics, the flux calibration had to be redone -- again using the ground-based
spectroscopic results of \citet{bal91} -- to determine the relevant exposure times.
With these exposure times in hand, all WFPC2 pixel brightnesses (from both Bally
mosaics and archival WFPC2 associations) have been accurately converted to absolute
fluxes/surface brightnesses, matching the ground-based flux calibration of \citet{bal91}.

\begin{figure}
\epsscale{1.0}
\centerline{\rotatebox{270}{\scalebox{1.0}[1.0]{
\plottwo{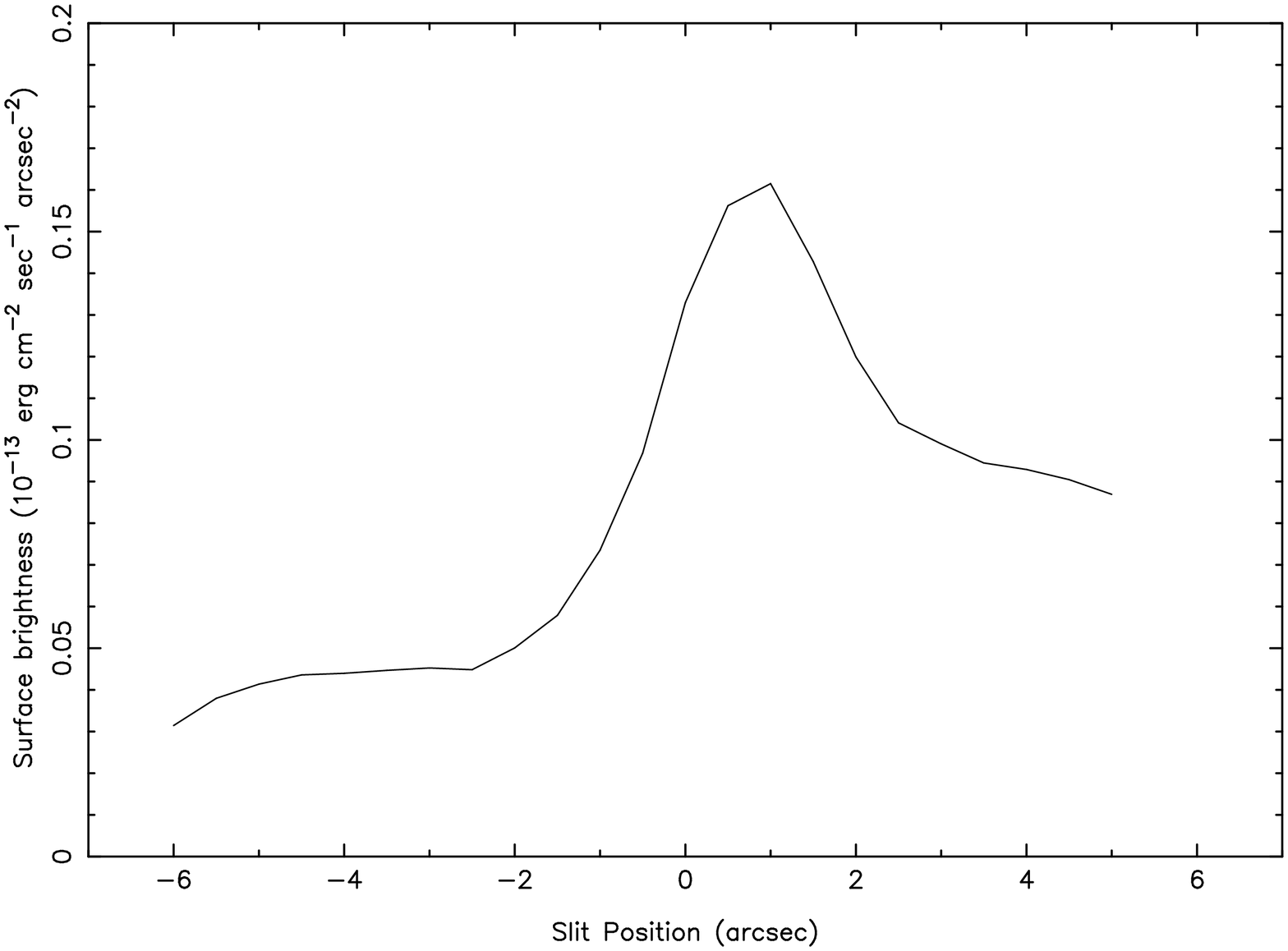}{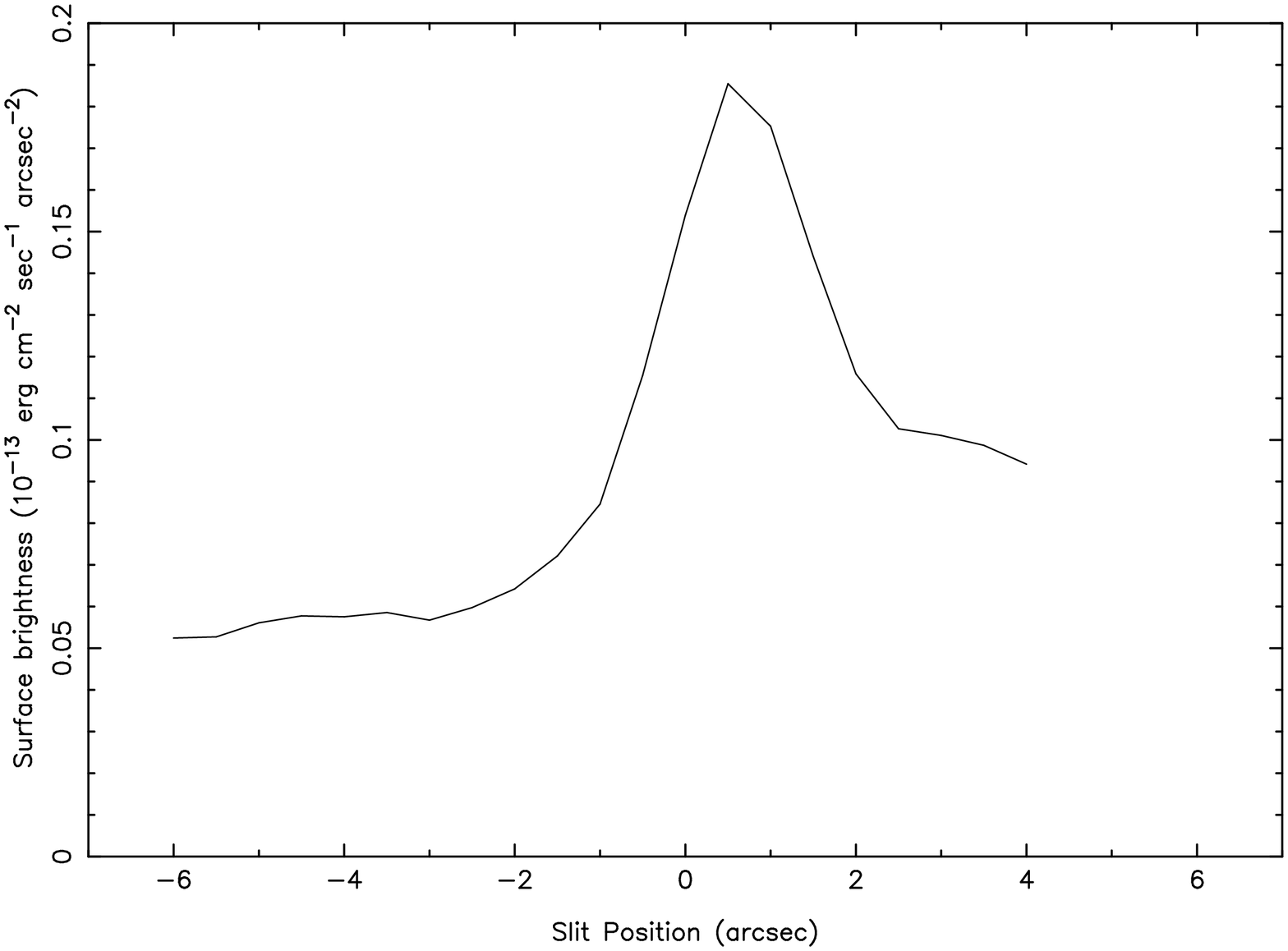}
}}}
\caption{
Flux variation of the He~{\sc i} 5876 spatially-narrow velocity-bridge component (+ 
continuum) along the slit in the blue spectrum (top) and in the red
spectrum (bottom).
\label{5876bridge} } \end{figure}

Looking at the spatially-resolved `x2' echelle spectra, we have noticed 
two distinguishing features associated with the shock feature: a 
wide ($5\arcsec$) velocity-shifted 
component and a narrow ($2\farcs5$) velocity bridge seemingly connecting 
the 
nebula and the shock.
Such a bridge
feature -- which can also be seen in Fig.~6 of BOM --
appears only to be associated with   
the leading optically visible shock.
As this feature is intrinsically narrow
($1\farcs5$ from WFPC2 images), we have 
been able to determine
the effective seeing
for the red and blue spectra
by measuring the
width (along the slit) of the He~{\sc i}~5876 bridge feature -- 
a line that is found in both the red and blue spectra
(see Fig.~\ref{5876bridge}).  The seeing was slightly different
on each of the two observing nights: $2\arcsec$ for the red observations and 
$2\farcs5$ for the blue.
We can also see from this figure
that the blue and red slits are aligned along their
lengths to an accuracy of $\sim0\farcs2$.  However, there
are small differences in the absolute observed flux, most
probably as a result of a position
difference in the transverse direction, along the shock feature.
These deviations will be 
addressed in \S~\ref{bluered}.

\begin{figure}
\epsscale{0.6}
\centerline{\rotatebox{270}{\scalebox{0.6}[0.6]{
\plotone{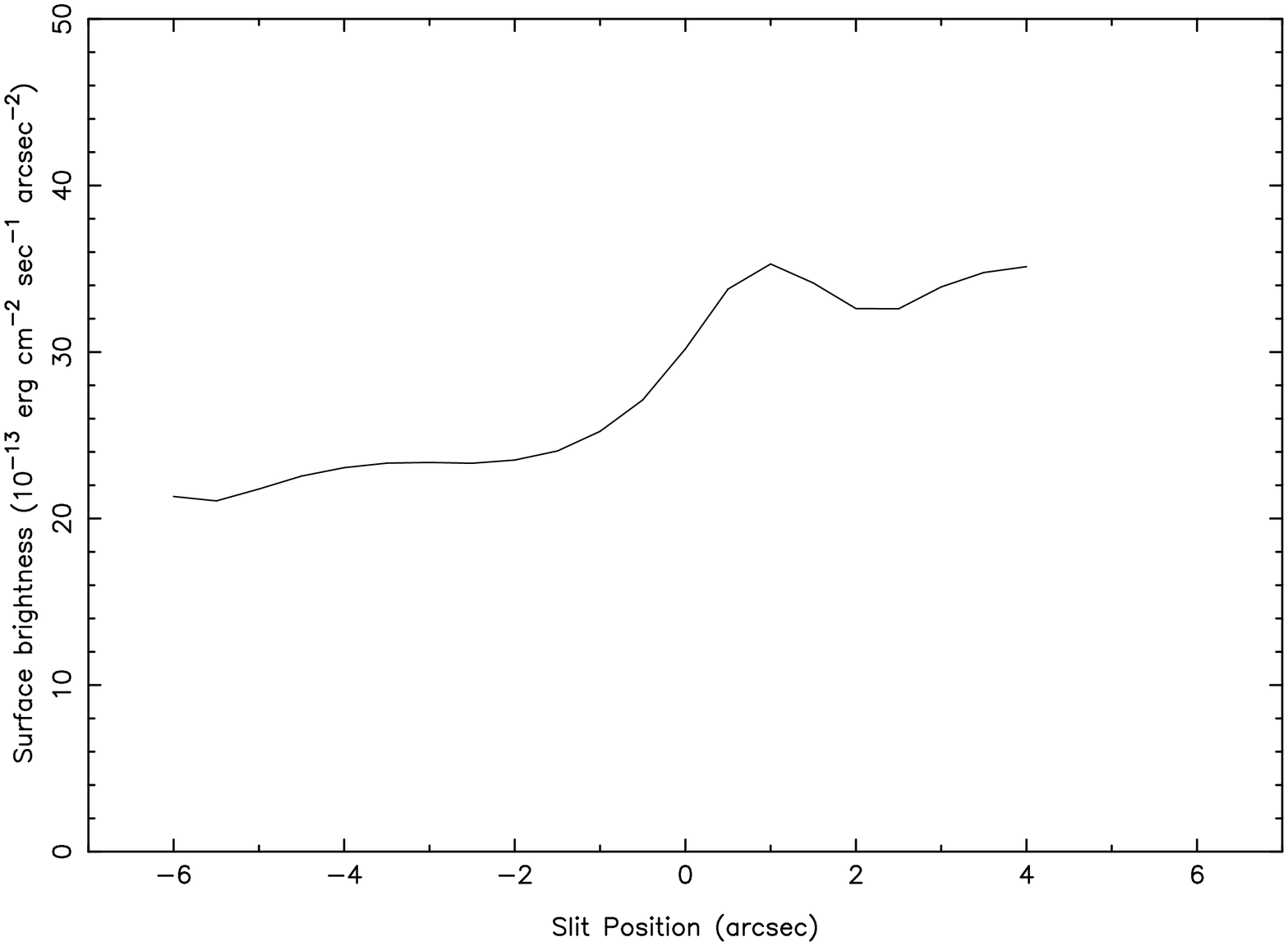}
\plotone{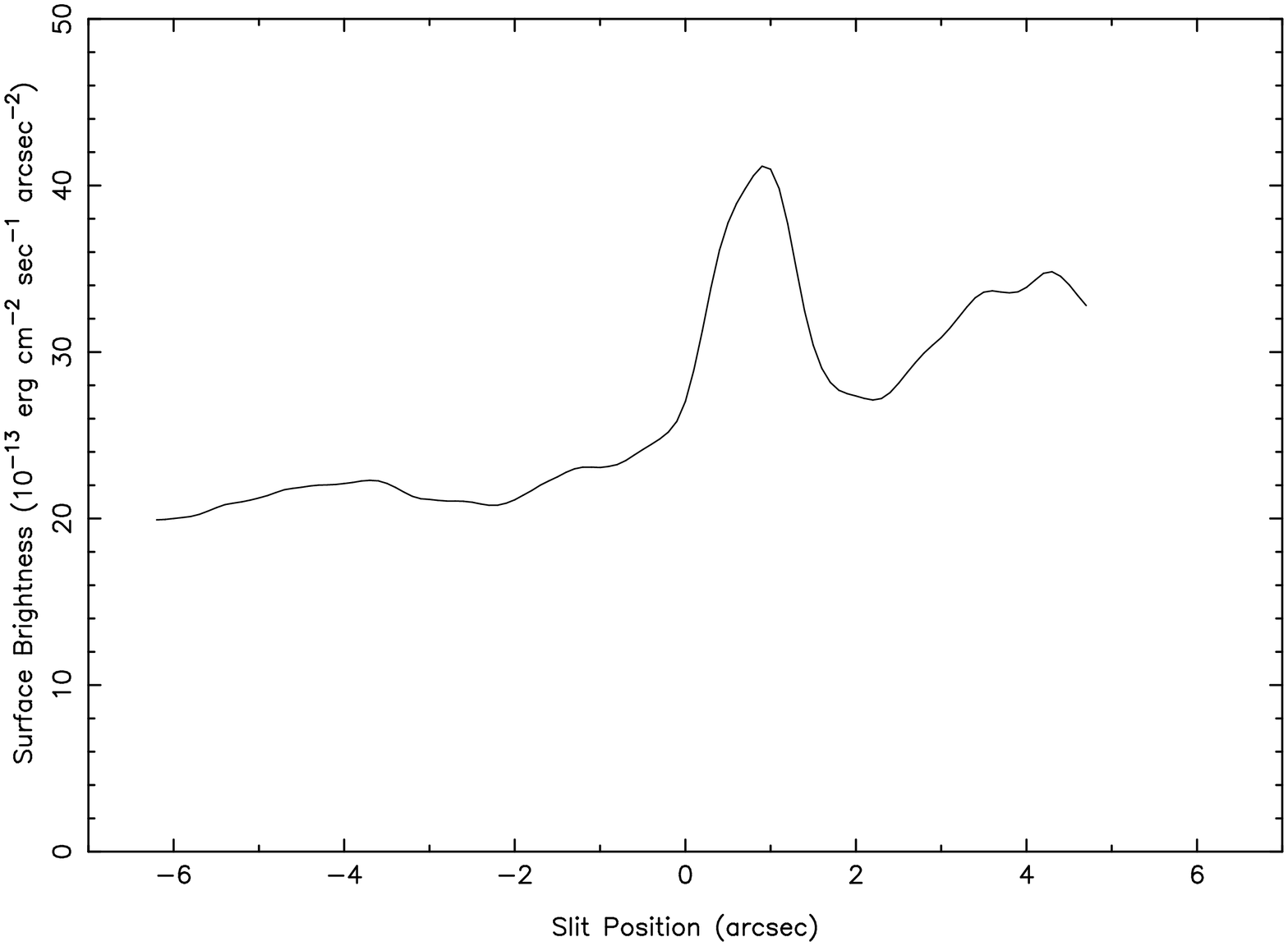}
\plotone{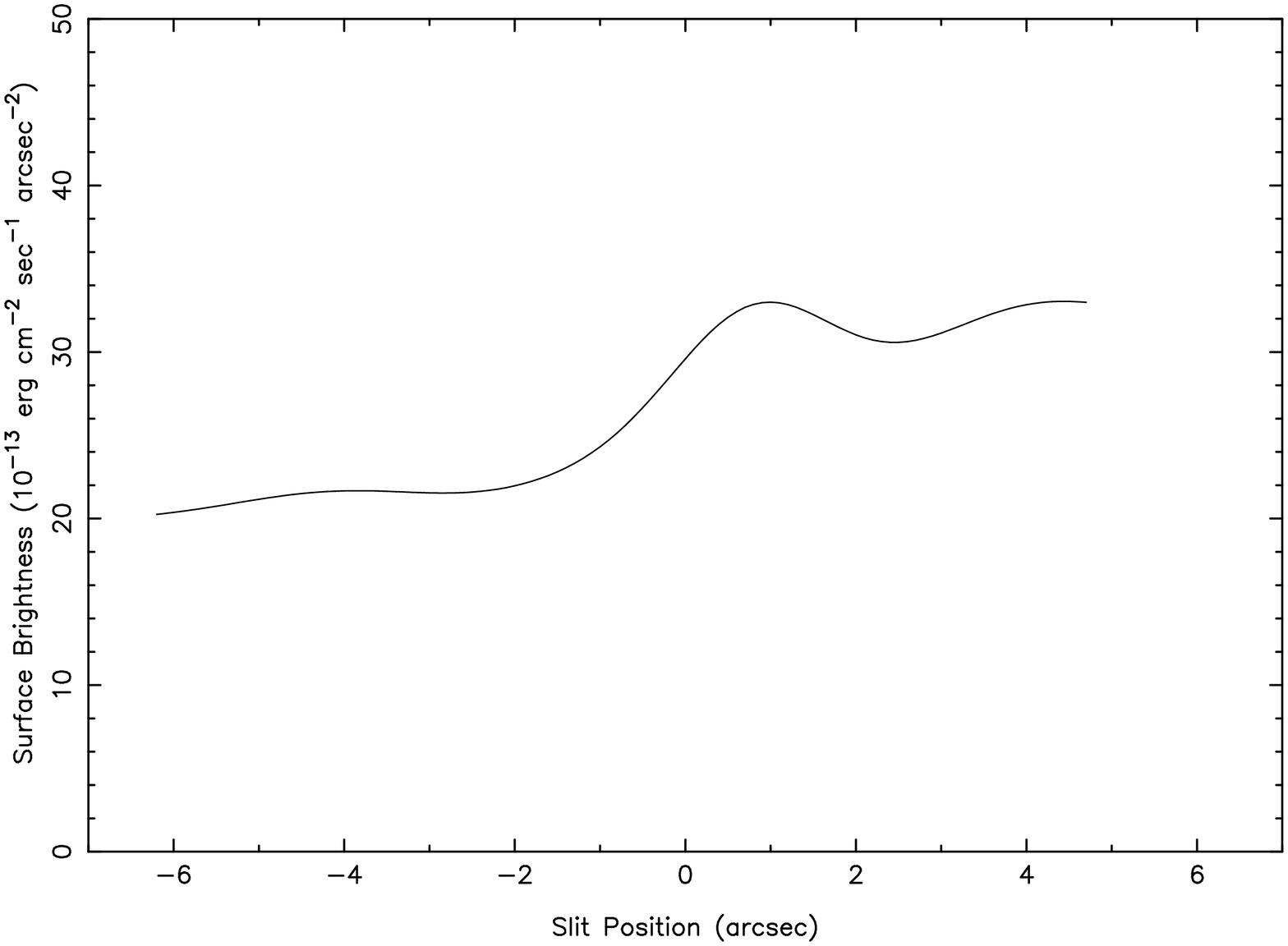}
}}
}
\caption{H$\alpha$ region comparison between ground-based echelle 
spectroscopy (top) and
WFPC2 photometry (F656N) (middle, bottom) as a function of slit position.
The bottom panel is the result of a $2\arcsec$ Gaussian convolution of
the WFPC2 slit extraction (middle) to simulate the $2\arcsec$ seeing.
These include H$\alpha$, continuum and line contamination from neighbouring lines
(namely, [N~{\sc ii}]).  \label{wfpc2slit}}
\end{figure}

A direct comparison of spatial variation in ground-based
and space-based observations over the same wavelength range was made to confirm
that the flux calibration of the echelle spectra is robust and
that the slit alignment and orientation are correct.
The echelle spectra were extracted over the same wavelength bandpass as the F656N
WFPC2 filter.
With knowledge of the approximate slit position (from a Polaroid of the slit
against the nebular background), the F656N flux-calibrated image was used to re-create the
expected spatial variation along the slit.
This re-created profile was
convolved with an appropriate-width
($2\arcsec$) Gaussian to simulate the seeing of the ground-based observations
(see Fig.~\ref{wfpc2slit}).
This processing allows for direct comparison between ground- and space-based
observations.
(Note that there has been no continuum (or line contamination)
subtraction from either the echelle spectrum or the WFPC2
reproduction, so the surface brightnesses in Fig.~\ref{wfpc2slit} are not those
of H$\alpha$.)
The slit's position on the F656N WFPC2 image was adjusted -- while maintaining
the slit orientation, $PA = 116^{\rm o}$ --  so as to emulate
more accurately the ground-based echelle slit spatial variation.
This 
required only a slight adjustment ($<1\arcsec$) of the slit from its original position on 
the WFPC2 image.
Using the slit position determined from this analysis, we compared all
WFPC2 filters with their respective portions of the ground-based spectra,
resulting in accurate reproductions of both the spatial variation and
absolute flux.

\begin{figure}
\epsscale{1.0}
\centerline{\rotatebox{270}{\scalebox{1.0}[1.0]{\plottwo{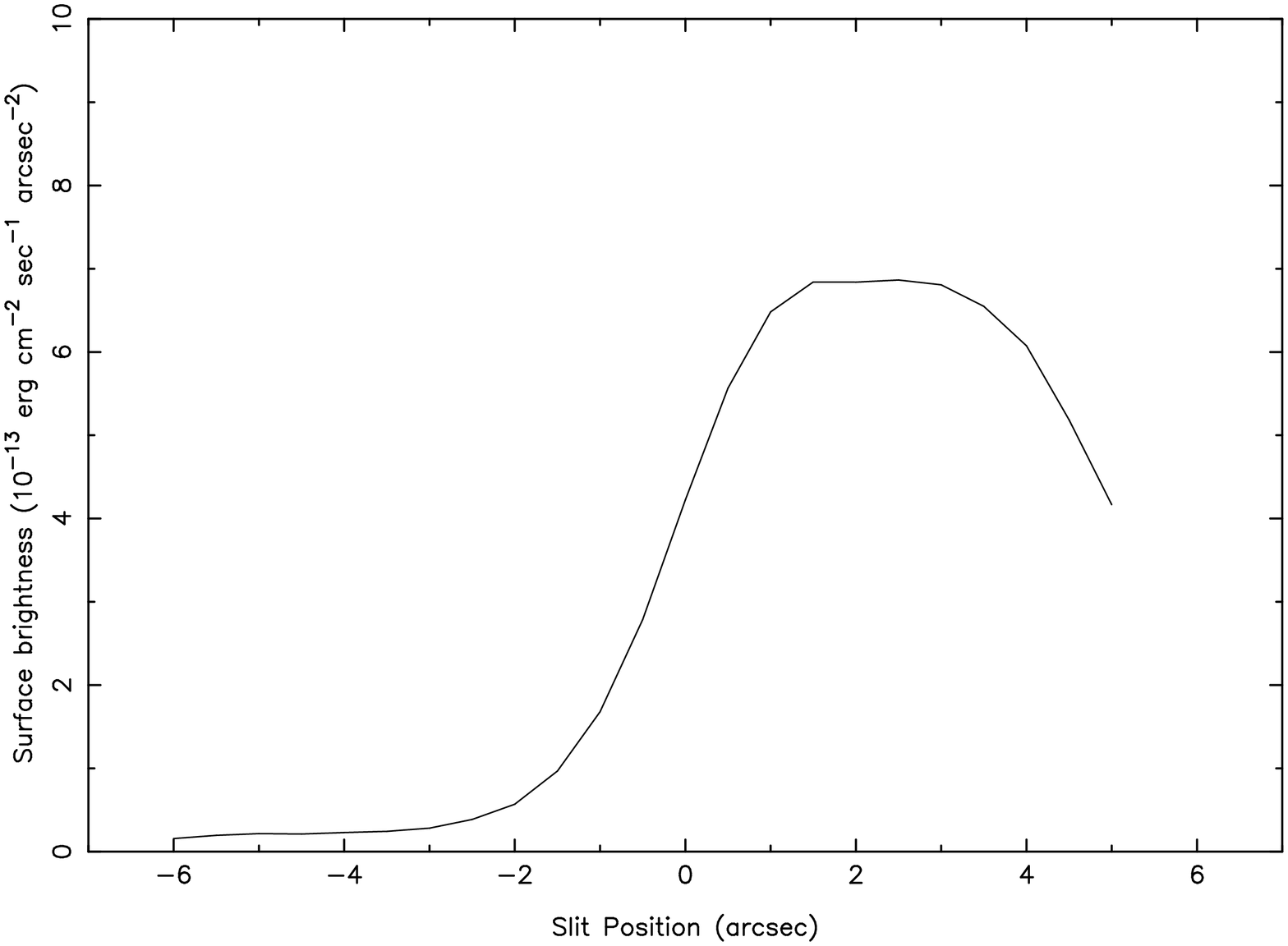}{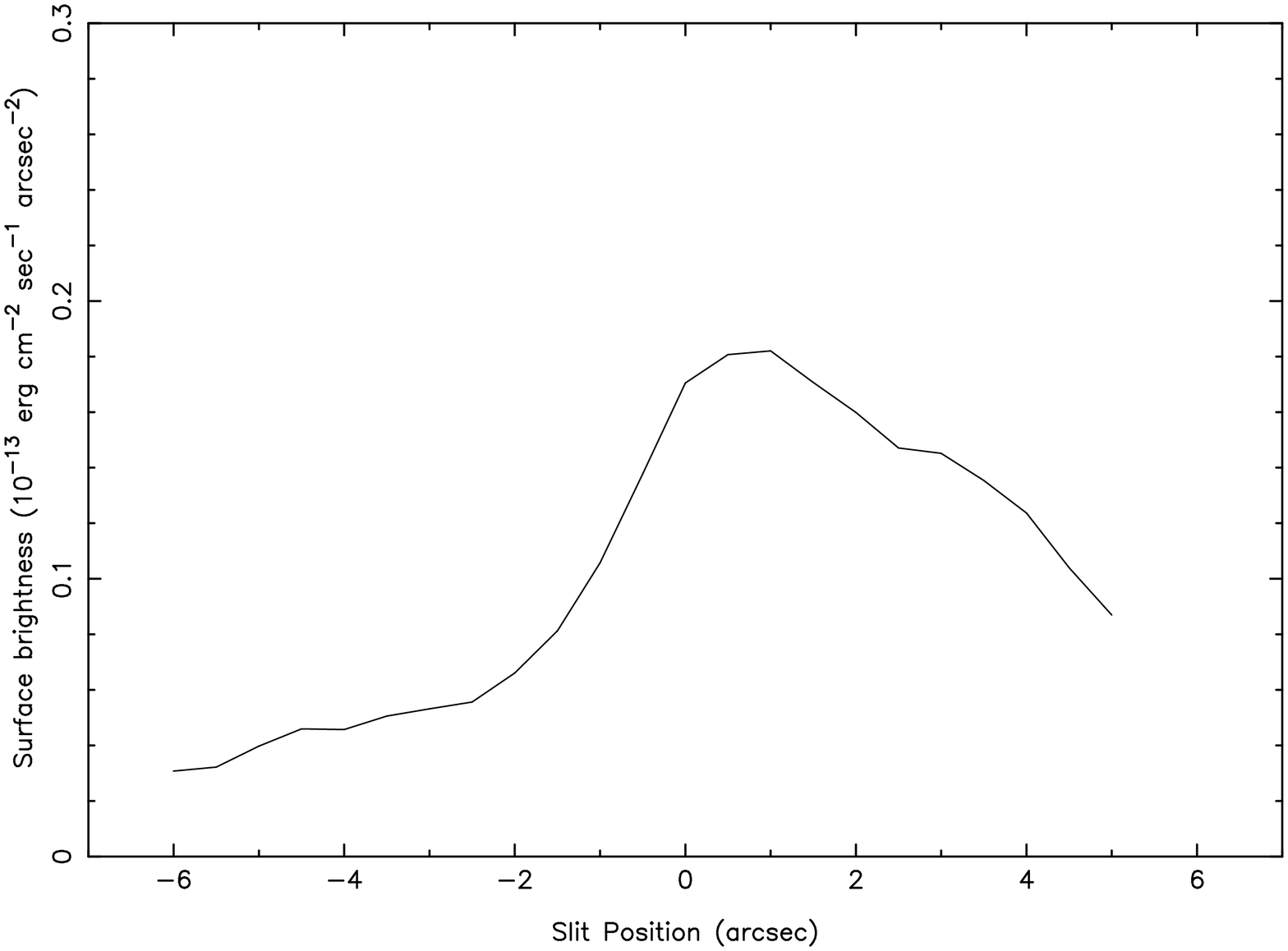}}}}
\caption{Variation of the [O~{\sc iii}] 5007 (top) and [O~{\sc ii}] 3726
(bottom) velocity-shifted shock component line flux across the $12.5\arcsec$ slit
(25 pixels).  The greater O+/O++ ratio near the eastern, leading 
edge, indicates a somewhat higher density there.
\label{oiii_slit}}
\end{figure}

The high-resolution echelle spectra allow us to analyze the spatial variation of the nebula
and shock separately -- offering insight not possible from the WFPC2 photometry.
For example, the slit variation of the [O~{\sc iii}]~5007 and
[O~{\sc ii}]~3726 shock fluxes are shown in Fig.~\ref{oiii_slit}. 
Differences in variation across the slit between these two ions may be 
indicative of a higher density at the eastern-most edge of the shock:
a higher density would lead to more recombinations and a slightly 
higher ionization fraction for O$^{+}$.

\begin{figure}
\epsscale{0.8}
\centerline{\rotatebox{270}{\scalebox{0.9}[0.9]{\plotone{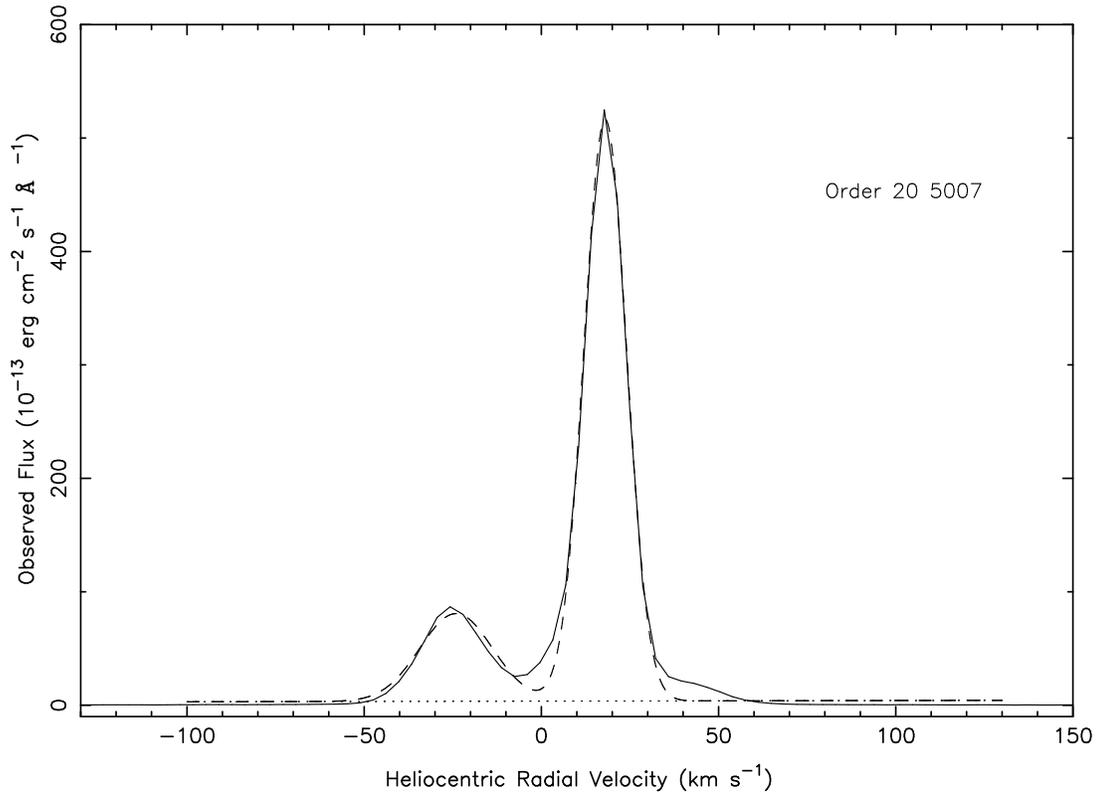}}}}
\caption{Double Gaussian fit of the nebular and velocity-shifted shock
components.  Eight parameters were used in the fit:
FWHM, peak wavelength and area for both components, and
two parameters to fit the continuum baseline level and
slope.
There is a third scattered light (red-shifted) component which
was not fit, explaining the poor fit redward of the nebular component.  The 
uncertainties
quoted in Table~\ref{redblue} reflect this poor fit.
The systemic (nebular) heliocentric velocity is $+18\pm2$~km~s$^{-1}$
for [O~{\sc iii}] \citep{ode01}.
\label{doublegauss}}
\end{figure}

These WFPC2 and shock component analyses suggest that
10 pixels ($-0\farcs5$ to $+4\farcs0$)
along the slit should be extracted in order to
obtain the best contrast between the background nebular component
and the velocity-shifted shock component
(referred to hereafter as simply the `nebular' and `shock' components).
Following this extraction, and with the nebular
line identifications from \citet{bal00} as a guide\footnote{
All ID wavelengths are from Atomic Line List v2.04 
(http://www.pa.uky.edu/\~{}peter/atomic/, maintained by 
P.~A.~M.~van~Hoof), except [O~{\sc ii}] \citep{bla04}.
}, the `x2' spectral features were fit 
with two 
Gaussian components representing the nebular and shock components,
as was done by \citet{doi04}. 
Eight parameters were used in the fit: FWHM, peak
wavelength and area for both components, and two parameters to fit the
continuum baseline level and slope.  The result of such a fit is shown in
Fig.~\ref{doublegauss}.

For cases where the shock component had a low
signal-to-noise ($S/N < 5.2$),
the lines were re-fit with a constrained double 
Gaussian.
The strong nebular component of the constrained fit had no 
constraints while the weaker shock
component's FWHM was fixed using the weighted average of the 
stronger lines' FWHM ($28.3$~km~s$^{-1}$).  The constrained velocity of the shock
component was set using the weighted mean of the 
H~{\sc i} shock components ($-42.1$~km~s$^{-1}$), and was maintained as a constant 
relative to the H~{\sc i} gas.
Because of the
ionization/velocity structure along the line-of-sight \citep{bal00},
the actual velocity 
differences between the weak shock component and the strong nebular
component depend on the ion.
If the $S/N$ of the shock component
improved and remained above $2.6$, the constrained fit was  
used.  Otherwise the double Gaussian fit was used for all lines with $S/N_{shock}
> 2.6$.

If the double and constrained Gaussian fits resulted in an undetectable 
shock component ($S/N < 2.6$), a five-component (FWHM, peak 
wavelength, area, continuum baseline and slope) single Gaussian fit was 
used for the nebular line.
The results of the line-fitting models are shown in Table~\ref{lines} with
nebular (neb) and shock (sh) components included in separate consecutive rows 
for each ID wavelength.
Column descriptions are included in the table

The shock component can be seen most prominently in the 
medium-ionization forbidden lines (e.g., [O~{\sc iii}]) as well as in the 
He~{\sc i} and H~{\sc i} permitted 
lines.  Although the shock component can also be seen in the 
low-ionization lines ([O~{\sc ii}], [N~{\sc ii}]), its 
strength relative to the nebular line is much weaker
(see Column~(9) of Table~\ref{lines}).
Of lines normally
associated with the ionization front (IF) of photoionized gas, some [S~{\sc ii}] can be seen 
very weakly in the shock component,
whereas others ([N~{\sc i}]) are too weak to be detected.
As will be discussed, the presence of [S~{\sc ii}] 
does not imply an ionization front in the shock.

Unfortunately, the [O~{\sc i}] sky lines\footnote{These lines are identified as such from     
sky spectra and other nebular spectra (at positions which did not
have a velocity-shifted feature) that were taken on the same evening.}
lie close to the wavelength where the shock component would be.
Using a triple Gaussian fit for the nebula, shock and sky components, we
can determine if there is a detectable shock component for the [O~{\sc i}]
6300 line.
The sky line FWHM, wavelength and area constraints are set by the sky line in
the 1SW echelle spectrum taken on the same evening;
the shock is constrained as in the constrained
double Gaussian case.
Following the fit of the three components in [O~{\sc i}] 6300, the shock component
has a null detection ($S/N\ll2.6$) lying well below our detection limit.
Neither is there a detectable bridge component as seen with the other shock lines.  It can
be safely said that [O~{\sc i}] (as with [N~{\sc i}]) line emission in the
shock lies below the detection limit for these spectra ($i.e., S/N < 2.6$).

At first sight this seemed at odds with the BOM HH~529 [O~{\sc i}] observations
depicted in their Figure 6 (WFPC2 631N image and Keck HIRES spectrum).
However, their detection of [O~{\sc i}] with the 631N filter
is not definitive due to contamination from
 the [S~{\sc iii}] line ($\lambda6312$) \citep{ode99}.
BOM's original HIRES spectrum shows a strong [S~{\sc iii}] velocity-shifted feature
($v\sim-39$~km~s$^{-1}$) associated with the eastern-most shock of HH~529
(O'Dell, private communication, 2005).  We also detect this in our spectrum and have
determined quantitatively that [S~{\sc iii}] would explain the presence of the shock in the
WFPC2 631N image.
Furthermore, the [O~{\sc i}] velocity contour plot displayed in BOM Fig.~6 is
actually an inadvertent copy
of the [O~{\sc iii}] plot (O'Dell, private communication, 2005).
The correct [O~{\sc i}]
contours are similar to the [S~{\sc ii}] contours in the west but have no velocity-shifted
feature in the east.

\subsection{Blue/red line strengths \label{bluered}}

Since the red and the blue spectra were taken on different nights, there
is a slight pointing uncertainty (see Fig.~{\ref{5876bridge}}) which makes 
comparison between the
red and blue spectra more difficult.
To study the uncertainties involved in inter-spectral comparison, we
identified lines that are found in both the red and blue spectra.
Six such lines had both a nebular and a measurable
shock component ([Fe~{\sc iii}]~5270,
[Cl~{\sc iii}]~5518, 
[Cl~{\sc iii}]~5538, Si~{\sc iii}~5740, [N~{\sc ii}]~5755, and
He~{\sc i}~5876).  Table~\ref{redblue} summarizes the results from the
(constrained) double Gaussian line-fitting for these seven lines
prior to applying the reddening correction.
The blue/red ratios for the nebular and
shock components are each shown separately
in Column~(8) of Table~\ref{redblue}, in the same rows as the
blue results.

The nebular lines measured from the blue spectrum
are not any stronger than the red on average ($B/R^{weighted}_{avg}\sim1.02\pm0.04$).
However, the average blue/red ratio ($B/R^{weighted}_{avg}\sim0.85\pm0.04$)
indicates otherwise for the shock.
This difference in blue/red ratios is not unexpected, as there is no 
reason to expect a correlation
between surface brightnesses in the nebula and shock.
Using these results, we make an
across-the-board
adjustment to all the red shock lines such
that the shock line strengths match between the red
and blue (0.85 adjustment) -- allowing for a complete (blue/red)
analysis of the shock.
No such correction is made to the nebular feature, whose blue/red ratio is 
consistent with 1.0.

\subsection{Reddening}
It is expected that the reddening of both the nebular and
shock components is the same, being dominated by foreground material.
However, prior to making the correction discussed in \S~\ref{bluered}, the
nebula and shock had drastically different H$\alpha$/H$\beta$
Balmer decrements: $4.99\pm0.04$ and $6.72\pm0.32$,
respectively.  After adjusting the line strengths so there is congruity  
between
the red and blue
lines (\S~\ref{bluered}) in the red and blue spectra and accounting for that 
uncertainty, these values
become $5.1\pm0.1$ and $5.7\pm0.4$ for the 
nebula and shock, 
respectively.  
This 
justifies
the use of the blue/red correction in \S~\ref{bluered} and the
use of the same reddening correction for both nebula and shock:  
$E_{B-V} =  0.3655$ \citep{mar06}.  The surface brightnesses are corrected 
for reddening 
as in \citet{mar06} and these dereddened values are included in Column~(7) of 
Table~\ref{lines}.

\section{Analysis}

\subsection{Velocity \label{analvel}}
Figure~\ref{shockvel} plots all the velocities determined from the shock
components of the Gaussian fits.  They are quite consistent, as expected
since unlike the expanding nebular gas,
there should be no velocity gradient in the shocked gas.
The shocked H~{\sc i} lines are shifted by $-42.1\pm1.2$~km~s$^{-1}$ 
relative to 
the nebular H~{\sc i} lines
(see Table~\ref{lines} and Fig.~\ref{shockvel}), or $-54.1\pm1.2$~km~s$^{-1}$ 
relative to the PDR 
in the molecular cloud, and hence, relative to the OOS embedded within the cloud.  
This agrees with the radial velocity measurements made 
by \citet{doi04} for the roughly coincident position 167-359 HH~529:
$-52$ to $-54$~km~s$^{-1}$ relative to the PDR/OMC-1.

The [Fe~{\sc iii}]~5270 shock component
(with $S/N\sim10$)
appears to
be discrepant in Fig.~\ref{shockvel}, with
velocities of 
$-32.9\pm1.1$ (blue) and $-31.9\pm1.1$~km~s$^{-1}$ (red).
This anomaly has an impact on the apparent nebular
velocity gradient of [Fe~{\sc iii}] lines
(see Fig.~10 in \citet{bal00}) and is taken up in Appendix~\ref{velgrad}.

\subsection{Temperature and density \label{tempdensity}}
Temperatures (in K) and densities (in cm$^{-3}$)
are calculated from emission line ratios using the NEBULAR routines
included within the iraf STSDAS package. 
These are summarized in Table~\ref{temden_nebshock},
with 
the respective
transition probabilities and collision strengths used in the calculations.

The $T_e$([O~{\sc iii}]) and $T_e$([N~{\sc ii}]) diagnostic 
lines can be seen in both the nebula and the shock, while the [O~{\sc i}]
temperature diagnostic lines can only be seen in the nebula.
The nebula temperature from the blue
[O~{\sc iii}] lines is
$T_e^{\rm neb}\mbox{([O~{\sc iii}])}\sim8536^{+35}_{-33}$,
whereas for the red [N~{\sc ii}] lines, 
the temperature is higher, $T_e^{\rm neb}\mbox{([N~{\sc ii}])}\sim10672^{+53}_{-52}$.
Although these temperatures come from the blue and red spectra respectively
and therefore represent two slightly different lines-of-sight, the temperature
rise with depth in the nebula is what is generally seen for other lines-of-sight,
and is largely the result of a hardening
of the radiation field as photons close to the ionization
limit are attenuated preferentially.
To complete the nebular temperature analysis, we have found
$T_e^{\rm neb}\mbox{([O~{\sc i}])}\sim8005^{+580}_{-408}$.

In the shock, the lines are weaker (in the case of [N~{\sc ii}], much weaker)
and therefore the calculated temperatures have much larger uncertainties.
The [O~{\sc iii}] temperature is $8366^{+252}_{-214}$, and that found from 
the [N~{\sc ii}]
temperature diagnostic lines is consistent (within $1\sigma$): $8784^{+1184}_{-729}$.
Since the shock is matter-bounded
(see \S~\ref{linestrength}), 
O$^{++}$ ([O~{\sc iii}]) and N$^{+}$ ([N~{\sc ii}]) are not
distinct zones and
the attenuation seen in the nebula is not possible.

The electron density can be calculated from the
diagnostic lines ([O~{\sc ii}] 3726, 3729; [S~{\sc ii}] 6716, 6731;
[Cl~{\sc iii}] 5517, 5537) which are seen in the nebula and weakly in the shock.
In the nebula, these three sets of density diagnostic lines cover slightly different 
ionization zones along a particular line-of-sight, but in the shock -- because of the
lack of distinct ionization zones -- the densities are expected to characterize the same
zone.
However, because of the disparity between red and blue slit positions, the calculated 
densities are also being defined along slightly different lines-of-sight.

For the nebula, we get
$N_e^{\rm neb}\mbox{[O~{\sc ii}]}\sim1939^{+50}_{-50}$ 
($N_e^{\rm neb}\mbox{[O~{\sc ii}]}\sim2164$ using entire slit) from the blue [O~{\sc ii}] 
lines.
The red [S~{\sc ii}] lines yield a much higher density,
$N_e^{\rm neb}\mbox{([S~{\sc ii}])}\sim5896^{+404}_{-366}$
($N_e^{\rm neb}\mbox{([S~{\sc ii}])}\sim5638$ using entire slit),
and the [Cl~{\sc iii}] lines yield an even higher density,
$N_e^{\rm neb}\mbox{([Cl~{\sc iii}])}\sim12074^{+1300}_{-1118}$.

It has been noted in \citet{est04} that the use of \citet{zei82} 
transition probabilities and \citet{pra76} collisions strengths drastically 
increases the calculated $N_e$([O~{\sc ii}]).  Upon further 
investigation, we find that a change in the transition 
probabilities alone will bring about the same result.  Using these older 
atomic data, we 
almost double the measured density: $N_e^{\rm neb}\mbox{([O~{\sc ii}])}\sim3811$,
bringing it more in line with the densities as measured from 
other indicators.
Another reason for questioning the atomic data comes from the [O~{\sc ii}] 
temperature 
-- which we overestimate slightly due to the 
shocked component impinging on the nebular component in the line pairs at 7320 and 
7330.  Using the density as 
calculated from [O~{\sc ii}] 3726/3729 (2000 cm$^{-3}$),
$T_e^{\rm neb}\mbox{([O~{\sc ii}])}\sim20000$~K.
However, with the larger density (4000 cm$^{-3}$) and the old atomic 
data, $T_e^{\rm neb}\mbox{([O~{\sc ii}])}\sim15000$~K.  An even larger density is required to 
reduce the temperature to 10000~K.  Note that these densities from [O~{\sc ii}] and
[S~{\sc ii}] are probably larger than in the more relevant [O~{\sc iii}] zone, because of a 
falloff of density in the expanding gas.
A similar result appears when we use older transition probability data for the 
$N_e$([Cl~{\sc iii}]) calculation.  The density is reduced to a more consistent value:
$N_e^{\rm neb}\mbox{([Cl~{\sc iii}])}\sim7247^{+575}_{-519}$.
To round out our discussion of density, we have looked at the density dependence of 
[Fe~{\sc iii}] (following \citet{kee01}) and O~{\sc ii} (following \citet{pei05}) lines.
The results are consistent with the densities we see in the rest of the nebula:
$N_e^{\rm neb}\mbox{([Fe~{\sc iii}])}\sim4700^{+800}_{-800}$ and 
$N_e^{\rm neb}\mbox{(O~{\sc ii})}\sim6700^{+100}_{-100}$.

The density of the shock is also calculated, but as the low-ionization
lines are weak, this calculated density is very uncertain. The blue [O~{\sc ii}]
lines yield $N_e^{\rm sh}\mbox{([O~{\sc ii}])}\sim2898^{+8429}_{-1997}$,
the red [S~{\sc ii}]
lines yield a density
near the limits of this diagnostic ratio, 
$N_e^{\rm sh}\mbox{([S~{\sc ii}])}\sim13183^{+10000}_{-11183}$,
[Cl~{\sc iii}] lines yield
$N_e^{\rm sh}\mbox{([Cl~{\sc iii}])}\sim21715^{+39170}_{-9641}$,
and the [Fe~{\sc iii}] lines yield\footnote{
The lower limit is set using [Fe~{\sc iii}]~4986 which is not observed in the shock.
This indicates that $I_{4986}$ is below the detection limit
($I_{\lambda}/I_{6678}\sim0.01$, or $I_{\lambda}/I_{4658}\sim0.05$), resulting
in a minimum density of 3200 \citep{kee01}.
}
$N_e^{\rm sh}\mbox{([Fe~{\sc iii}])}\sim7300^{+8000}_{-4100}$.
(The O~{\sc ii} lines are too weak to yield a consistent estimate
of the shock density.)
Use of the older atomic data again
results in a higher [O~{\sc ii}] density, $N_e^{\rm sh}\mbox{([O~{\sc ii}])}\sim7304$,
and a lower [Cl~{\sc iii}] density, $N_e^{\rm sh}\mbox{([Cl~{\sc iii}])}\sim10911$.
The shock density appears to be larger (by roughly a factor of two) than that of the nebula,
but given the large uncertainties, a density identical to that
of the nebula is also allowed by the line ratios.  Density will be revisited
in a discussion of
shock models in \S~\ref{anal.model}.

\subsection{Relative line strengths and ionization structure \label{linestrength}}

To maximize the shock-to-nebula ratio, the echelle spectra were extracted over only 
half the slit.
Even then, the echelle spectra
maintain a weaker shock component as compared to the nebular component (see Column~(9) of 
Table~\ref{lines}),
indicative of a lower
density, or more probably, a shorter emitting column in the shock.
Since the illumination of the shock is roughly the same as that of the nebula, if the
shock were optically thick,
the shock-to-nebula
ratio would be close to one for all lines, barring minor changes due to differences in
density (near the critical density) or changes due to abundance (see \S~\ref{cloudyx2_discussion}).
Here, the shock-to-nebula ratio is clearly lower than one,
and so the shock is matter-bounded.

The relative strength
varies from $0.2$ for the medium-ionization lines (e.g., [O~{\sc iii}]) to 
less than $0.03$
for the low-ionization lines (e.g., [N~{\sc ii}]) to below the detection
limit for the lines usually
associated with the ionization front (e.g., [N~{\sc i}])
and is plotted as a function of ionization potential 
in Fig.~\ref{ratio}.
In the case of a shortened emitting column,
the ionization potential serves as an indicator of
ionization fraction 
(where higher ionization potential
indicates higher ionization fraction)
while the shock-to-nebula ratio is a measure of the
optical thickness of the shock
to the relevant ionizing radiation.
H~{\sc i} is presented as a standard for shock/nebula ionization comparison
as its
originating ion (H$^{+}$) has an ionization fraction of roughly one
throughout
both the shock and the nebula.
The ratios of the medium-ionization species
([O~{\sc iii}], [Ar~{\sc iii}], [Ne~{\sc iii}]) all lie
above H~{\sc i} as they have a higher net ionization fraction in the shock than
in the nebula column.
However, none of these ratios is unity either.
Thus, for example,
in the shock there is not a complete O$^{++}$ zone, preceding a distinct
O$^{+}$ zone.
The ratios of the low-ionization species
([O~{\sc ii}], [N~{\sc ii}], [S~{\sc ii}]) lie below H~{\sc i} as they have a
lower ionization fraction in the shock than in the nebula.
In fact, they must arise from trace ionization stages in a more highly ionized zone (e.g.,
trace O$^{+}$ in the O$^{++}$ zone).  This is in contrast
to the nebular column in which  
lines arise from distinct ionization zones.
The lack of an ionization front tracer ([N~{\sc i}]) in 
the shock component provides further corroboration for a
matter-bounded shock.  

The critical densities associated with the [O~{\sc ii}], [S~{\sc ii}] and 
[Cl~{\sc iii}] line transitions need to be considered as these lie within 
the expected density range of the shock and so collisional de-excitation
could contribute to the 
relative weakness of the shock lines.  However, the weak [N~{\sc ii}] 
lines have critical densities of $7.8\times10^4$ and 
$1.2\times10^7$~cm$^{-3}$ which lie well above the model-predicted density 
as discussed in \S~\ref{anal.model}.
The predominant cause of weakness is the lack of parent ions in this highly-ionized 
matter-bounded geometry.

\subsection{Temperature fluctuations \label{profiles}}

Temperature fluctuations ($t^2$),
first defined/introduced by
\citet{pei67},
have been popular in explaining the
differences in abundances found from forbidden lines as compared to those
found from permitted lines.
Although these fluctuations have been deduced to exist, 
their deduced size ($t^2\sim0.02$)  has not been explained.
\citet{fer01} has suggested a possible link with additional photoelectric
heating from grains.
Other suggestions -- large scale variations in $T_e$, or the
presence of regions either shielded from direct illumination by $\theta^1$ Ori C
or heated by shocks (from SNe mainly) -- might explain temperature fluctuations in 
the nebula, but not in a small-scale shock.

O~{\sc ii} permitted and [O~{\sc iii}] forbidden lines can be used to 
infer a value of $t^2$ as has been done by \citet{est98} and 
\citet{est04} for the nebula.  We apply this to the shock too, adjusting our
permitted line analysis to allow for deviations from LTE \citep{pei05}.
First, we must confirm that the nebular and shock O~{\sc ii} permitted lines
form following recombination \citep{gra76}.
The shock-to-nebula ratios of the O~{\sc ii} and
[O~{\sc iii}] lines are the same, and
much larger than the shock-to-nebula ratios of the
[O~{\sc ii}] lines.
Also, note that the velocities of the O~{\sc ii} lines are consistent with the velocities
of [O~{\sc iii}] in the nebula (Table~\ref{lines}).  
  These two observations both
confirm that the O~{\sc ii} lines are actually a result of recombinations 
from O$^{++}$ and not a result of direct starlight excitation of O$^{+}$,
validating the use of these lines in the determination of the 
O$^{++}$/H$^+$ ratio.  We have
used O~{\sc ii} recombination line multiplet~1 and [O~{\sc iii}] collisionally-excited
lines 4363, 4959 and 5007 with the NEBULAR\footnote{
The collisionally-excited line
results were calculated using the three-zone model in IRAF.
In this case only the low- and medium-ionization zones (those of O$^{0}$/O${+}$ and O$^{++}$) 
are of interest.
The adopted densities of the nebula and shock are $N_e = 6000$ and $10000$, respectively.
The temperatures are those determined
from the [N~{\sc ii}] and [O~{\sc iii}] temperature diagnostic lines (refer to 
\S~\ref{tempdensity}) for the low- and medium-ionization zones, respectively.
} routines in IRAF (as in \citet{est98,est04})
to determine $t^2$ for the nebula and the shock.  Not all permitted lines of 
O~{\sc ii} multiplet~1 are observed, so individual (or pairings of) recombination lines
are used to predict the complete multiplet's relative surface brightness (see 
Table~\ref{tempfluct}), following \citet{pei05} (their equations 3 and 4).
Using case A and case B O~{\sc ii} recombination coefficients from 
\citet{sto94} and case B H~{\sc i} recombination coefficients from \citet{sto95},
O$^{++}$/H$^{+}$ is calculated (see Table~\ref{tempfluct}).

The O$^{++}$/H$^{+}$ abundances from recombination and 
collisionally-excited lines and the inferred $t^2$ are summarized in 
Table~\ref{RLCEL} for both the nebula and the shock (along with the O$^{+}$/H$^{+}$, O$^{0}$/H$^{+}$
and total O/H abundances).
Our nebular $t^2$, $0.009\pm0.004$, is much lower than what
has been deduced from another line-of-sight (for the same O$^{++}$ ion), 
$t^2\sim0.020\pm0.002$ \citep{est04} -- which did not correct O~{\sc ii} lines for deviations 
from LTE.
Despite the presence of detectable O~{\sc ii}
lines in the shock, the uncertainties are large enough that there is only 
a $1\sigma$ ``detection'' of $t^2$ in the shock, $t^2 = 0.010\pm0.010$.
If the grains are depleted in the shock, a detectable $t^2$
suggests that the grains may not be the main contributor to $t^2$.  This will be 
followed up in \S~\ref{tempfluc2}.

\section{Models \label{anal.model}}

The HH object has been shown to be photoionized, so we can model
the emission using the radiative-collisional equilibrium code, Cloudy.  
As the [Fe~{\sc iii}] lines figure prominently in our discussion, we have improved the description of the
Fe$^{++}$ atom in Cloudy from a two-level to a 14-level atom, using collision 
strengths and transition probabilities from
\citet{zha96} and \citet{qui96}
respectively. This allows  all multiplet 
lines associated with $\lambda4658$ and $\lambda5270$
to be included in the determination of Fe abundance.
Also, as the accuracy of the atomic data for O$^{+}$ has been questioned 
(\S~\ref{tempdensity}, \citet{est04}) we have 
replaced the up-to-date transition probabilities \citep{wie96} with the 
older ones \citep{zei82}.

\citet{bal00} showed that the incident continuum radiation (from the ionizing star, $\theta^1$~Ori~C) is 
best represented by a Mihalas stellar atmosphere model.  However, to test the robustness 
of our result, we also developed models using a Kurucz stellar atmosphere.  Note that
the issues with the Kurucz atmosphere (primarily with its inability to accurately
predict the high ionization line [Ne~{\sc iii}]~3869)
are not that relevant to our discussion of low- and medium-ionization species.

Since the shock has
a 
small
covering factor compared to the nebula,
spherical geometry is not assumed and
an inner radius is not set.
The sound-crossing time 
for the HH feature ($\sim10^3$ years) is roughly the same order as the 
dynamical timescale of the flow ($1500$ years), so instead of assuming a 
constant pressure
(as would be the case in a nebular model),
we assume a constant density.
Also, as the flow has only been in existence for
1500 years ($5\times10^{10}$ s), it is important to check the 
validity of a photoionization equilibrium code.  The
longest timescale from the Cloudy shock model comes from 
H-recombination: $2\times10^8$s -- well within the limit of the flow's age.
The incident surface flux of ionizing photons, $\phi(H)$ should be close to 
the value derived
for nebular models (log$\phi(H)\sim13.0$, $e.g.$~\citet{bal91})
as the shock
is roughly the same distance from $\theta^1$~Ori~C as the nebula (see \S~\ref{intro}).
However, the electron density 
is probably significantly higher in the shock than in the nebula as evident from the 
observed $\lambda6731/\lambda6716$ ratios.
Since the shock has been shown to be matter-bounded and
homogeneous with respect to its ionization structure
(\S~\ref{linestrength}), the shock model can be developed simply as a finite thickness
truncated nebula (i.e., with a pre-defined stopping thickness).
This thickness can be predicted
from the length ($10\arcsec$) and width ($2\arcsec$) of the shock in
the plane of the sky (from [O~{\sc iii}] WFPC2 image) and its assumed  
cylindrically-symmetric geometry.  Adopting a distance to the nebula of
460 pc (BOM), the predicted median depth ($3\arcsec$)
translates to a thickness of 0.007 pc ($2\times10^{16}$ cm). 

The parameters are varied from these initial values, using observed surface
brightness of He~{\sc i}~6678, and line ratios indicating temperature, 
density and ionization (see Table~\ref{constraint}) to determine the best-fit
models.
In the case of an
optically thin model, the surface brightness varies as $n_H^2 t$, where $n_H$ is  
the hydrogen density and $t$ is the model thickness.  Adjusting the model thickness
does not result in (much of) a change to any of the other constraint ratios as the
ionization fractions of most species are constant through the entire model.
Therefore, $t$ is not completely independent, leaving
$T_{\star}$,
$\phi(H)$ and $n_H$ as the three independent parameters.

A series of models were developed, two of which are summarized in Table~\ref{cloudy}:
one with a Mihalas stellar atmosphere and Cloudy Orion abundances 
(from \citet{bal91,rub91,ost92});
and one with a Kurucz stellar atmosphere and \citet{est04} Orion abundances 
(see Table~\ref{abund_table}).
After determining the best-fit parameters for
both of these models, the Fe abundance was adjusted to fit
the series of [Fe~{\sc iii}] lines using the Cloudy $optimize$ routine.  Some 
implications
of the derived abundances will be discussed in \S~\ref{depletion}.

\section{Discussion \label{cloudyx2_discussion}}

The echelle observations (from Table~\ref{lines}) and
the model predictions
are summarized in
Table~\ref{modeltab}
as $I_{\lambda}/I_{6678}$.  If there is no model prediction (i.e., the particulars of
the line formation are not included in the model) then the
observations are not included in the table.

It is informative to compare the model predictions with the echelle
observations for not only the constraint ratios, but
all
lines predicted by the model.  This will further test the robustness of the model.
Special note
should be taken of lines predicted to be seen in
the 
shock,
but not observed.
Of such cases,
many of them appear around or below the detection limit
($I_{\lambda}/I_{6678}\sim0.01$).
Many of those lines predicted to be above this limit 
(He~{\sc i}~3705, [S~{\sc iii}]~3722, H~{\sc i}~3722, He~{\sc i}~3889, He~{\sc i}~4009,
[S~{\sc ii}]~4076, C~{\sc ii}~4267, O~{\sc ii}~4341,
[O~{\sc ii}]~7320, [O~{\sc ii}]~7331) appear as blended line
features in the spectrum
and therefore are not included in Table~\ref{modeltab}.
There are another three undetected-but-predicted shock lines:
O~{\sc ii}~4093,
O~{\sc ii}~4111,
O~{\sc ii}~4277.
Each of these is a complete multiplet prediction requiring a series of multiplet
correction factors to predict the observed multiplet component lines.
After applying these correction factors to the shock model lines, 
their predicted flux would lie below the observed detection limit.
As discussed in \S~\ref{observe}, the velocity-shifted [O~{\sc i}] lines are
sky lines and not associated with the shock, explaining the disagreement
between observation and model at [O~{\sc i}]~6300.

\subsection{Depletion \label{depletion}}

The Orion nebula is
thought to have a depleted gas-phase abundance of Fe of roughly a
factor of 10 (with respect to solar) due to the presence of grains.
From a preliminary analysis, this does not appear to be the case for the 
shock.
The ionization fraction of Fe$^{++}$ remains roughly constant through the slab
(Fe$^{++}\sim0.2$, Figure~\ref{ionizeiron}) with no well-defined
Fe$^{++}$ zone, and yet the 
[Fe~{\sc iii}]
lines appear quite strong relative to the nebula lines (see 
Figure~\ref{ratio}).
This may indicate an ``undepletion'' of Fe (possibly up to the solar level).

A series of [Fe~{\sc iii}] lines ($\lambda4658$, $\lambda5270$, etc.)\ is predicted
using the higher resolution Fe$^{++}$ ion (\S~\ref{anal.model})
and numerous [Fe~{\sc ii}] lines are predicted
using the 371-level Fe$^{+}$ ion.  These [Fe~{\sc ii}] lines have been shown to
have large contributions from continuum pumping \citet{ver00}
and therefore, cannot be used as indicators of Fe abundance, but the modelled [Fe~{\sc iii}]
lines scale linearly with the Fe abundance.
The iron abundances determined from matching the observed and modelled [Fe~{\sc iii}]         
lines in both
shock models appear to be roughly consistent with the nebular gas-phase Fe abundance
(see Table~\ref{depletion_table})
indicating that
the seemingly high shock [Fe~{\sc iii}] line strengths can mostly be
explained by differences in the
models' parameters,
not needing to resort to an order of magnitude change in the abundance.  However, 
if the nebular Fe/H gas-phase abundance is as low as 6.23 \citep{est04}, the extreme
prediction of Model B would suggest a three-fold increase in Fe/H gas-phase abundance indicating
a partial destruction of grains in the shock.

An analysis of the Fe abundance of Orion B stars \citep{cun94} and a follow-up analysis
of Orion F
and G stars \citep{cun98} imply that the total abundance of Fe is consistent from 
star to star within the
Orion association, but that there may be a slight total Fe depletion 
with respect to solar (-0.16~dex, \citet{cun98}).
The Fe depletions obtained from our shock analyses are greater, ranging
 from -0.8 to -1.0 dex with respect to solar 
-- on 
the order of the depletions
found in the nebula \citep{bal91, rub97, est98, est04}.  Assuming that the
total Orion Fe abundance is on the order of that found from the Orion association
stars, 
the majority of the iron in the shock, as in the nebula, must be locked up in grains.

A number of Si lines are also seen in the shock.  
Although there is no Cloudy prediction for these Si lines, the 
observations can still be analyzed using ionization models from Cloudy and 
line information from \citet{gra76}.
The shock-to-nebula ratio is high ($\sim0.15$) for Si~{\sc 
ii}~3856,
5056,
6347
(and 6371), but these
lines have been shown to form due to starlight 
excitation \citep{gra76} in the Si$^+$ gas.
The Si$^{+}$ ionization fraction predicted from the Cloudy models (0.03)
is much less than that for Fe$^{++}$ (0.2), but the Si~{\sc ii} lines
are not linearly dependent on Si abundance so these lines alone can not
be used to determine Si abundance.

Since $\sim20\%$ of O atoms are thought to be in dust grains \citep{est04},
the gas-phase abundance of O can be analyzed to determine the extent of dust destruction.
The total O/H in the nebula and in the shock is summarized in Table~\ref{RLCEL}.
Note that O/H for the shock component ($8.73\pm0.05$) is an upper limit and the actual value is most
likely closer to that of O$^{++}$/H$^{+}$ ($8.69\pm0.05$).
The shock [O~{\sc ii}] and [O~{\sc iii}] line profiles across the extracted part of the slit
peak at different spatial positions (see Fig.~\ref{oiii_slit}),
indicating that these lines are tracing physically different
lines of sight and that a simple addition of O$^{++}$ and O$^{+}$ may overestimate the O/H abundance.
Our observed nebula O/H abundance ($8.48\pm0.01$ or $8.52\pm0.03$ using recombination lines)
deviates slightly from other Orion
nebula observations, which
find O/H$\sim8.60-8.65$ \citep{bal91,rub91,ost92,est04}.
The shock O/H abundance should be compared to an average/typical O/H nebula abundance, as the 
shock originates in a different region of the nebula.
For our observations of the shock, the uncertainty in O/H is large enough that no definitive
statement can be made with regards to dust destruction in the shock, except that there
may be a small ``undepletion'' of gas-phase O to parallel the ``undepletion'' of gas-phase Fe.

\citet{smi05} have imaged the bow shocks of HH~529 with T-ReCS at $11.7\mu$, seeing what they
refer to as ``most likely thermal dust emission'' associated with the eastern-most shock.
Although supporting the argument of \citet{smi05},
our evidence for the existence of grains in this one HH object is
anomalous when compared with the 21 HH objects studied by
\citet{boh01}.  For both their high-excitation/fast-moving ($v~>~85$~km~s$^{-1}$)
and low-excitation/slow-moving ($v~\leq~50$~km~s$^{-1}$)
HH objects, the derived Fe depletion is never more than $-0.4$~dex
suggesting that the grains are most likely destroyed in the HH objects regardless
of their velocity.  It is of interest that for HH~529 -- measured to have a velocity 
of
$76$~km~s$^{-1}$ relative to OMC-1 -- the depletion is on the order of that of
the nebula ($-1.0$~dex);
there is no evidence for the complete
destruction of grains in the eastern-most visible
shock of HH~529.  This is more along the lines
of what one would expect: a slow-moving flow would not be expected
to destroy grains, whereas a fast-moving flow would.  \citet{boh01} suggest
that the molecular cloud material currently associated with
their slow-moving shock may have had its grains destroyed in 
an earlier pass through a faster-moving shock.
Following this argument,
the material associated with HH~529 must
not have ever passed through a high-excitation/fast-moving shock.
This is slightly inconsistent with the set of HH~529 velocities
measured by \citet{doi02,doi04}, many of which suggest
the material may have been travelling faster than 85~km~s$^{-1}$.
A full Fe abundance analysis of all HH~529 shocks could offer further
insight into grain destruction in Herbig Haro objects.

\subsection{Temperature fluctuations \label{tempfluc2}}
The $t^2$ deduced to exist in the nebula is $0.009\pm0.004$
and in the shock is $0.010\pm0.010$
(\S~\ref{profiles}) (which are both within $2\sigma$ of zero).
Two suggested explanations for the existence of $t^2$ --
large scale variations in $T_e$ or the presence of shielded or
heated regions -- can not apply for the small column
covered by the shock.
However, since the grains still appear to be present in the shock
a $t^2$ detection suggests a third explanation: that the grains may be the 
main contributor to $t^2$.
Conversely, an ``effective'' $t^2$ may 
be introduced if the
effective recombination coefficients, collision strengths and/or transition probabilities are
inaccurate, or if there were
some other contributions to the line emission besides solely
recombination or collisional excitation.

\section{Conclusions \label{conclusions}}

High-resolution spectroscopy of the Orion nebula across the 
Herbig Haro object HH~529 has allowed for a comparison of that local part 
of the nebula with the velocity-shifted spectrum of the flow.  The radial
velocity (as measured from the H~{\sc i} emission lines), $-42.1\pm1.2$~km~s$^{-1}$ 
is consistent with the $-40$  to $-42$~km~s$^{-1}$ range
as measured by \citet{doi04} for 
a slightly different line-of-sight.  In addition, there is ample evidence 
to suggest that this flow has been photoionized.  Herbig Haro objects 
usually have a strong low-ionization line spectrum.  In this case, the fact 
that we see strong medium-ionization lines and much weaker low-ionization 
lines indicates that we have a photoionized shock, as first suggested 
by \citet{ode97a}.  The distinguishing shock-to-nebula ratios as a function 
of ionization fraction, or ionizing potential as in Figure~\ref{ratio}, 
further support this hypothesis, leading us to model the shock as a 
matter-bounded photoionization region.

The
shock component was modelled using the 
photoionization equilibrium code, Cloudy.
Both Mihalas and Kurucz stellar atmosphere models were investigated to ensure
the robustness of our conclusions.
A series of ``best-fit'' models covering a range of stellar temperatures,
densities, and $\phi(H)$ fluxes has allowed us to determine that the 
depletion
of Fe (relative to solar) in the nebula 
also exists in the shock.
The higher density of the photoionized shock allows for the formation of
relatively strong [Fe~{\sc iii}] lines without necessitating a reduction
of the Fe depletion.  The Fe depletion for the shock is roughly the same as for the 
Orion nebula, an order of magnitude relative to solar (-1.0 dex).
The total Fe abundance of the Orion association stars may be slightly depleted
(-0.16 dex, \citet{cun98}), but not to the extent of the gas-phase Fe in the 
nebula and shock.
This suggests that if the total Fe abundance in the nebula and shock is
of the same order as that found from the Orion association stars, grains 
must be present 
in the Herbig Haro flow to account for the depletion of gas-phase Fe.
\citet{boh01} suggests that grains are destroyed in many HH objects as the material
passes through high-excitation/fast-moving shocks.  From our results,
we infer that the 
eastern-most shock of
HH~529 never reached the velocities necessary to destroy the majority of the
grains despite the presence of fast-moving shocks elsewhere in HH~529.
This supports the observations of 11.7$\mu$ thermal dust emission in the eastern-most
shock of HH~529 \citep{smi05}.
Further information about grain destruction in HH~529
can be obtained from parallel Fe abundance analyses for the remainder
of the HH~529 photoionized shocks.

Temperature fluctuations in the Orion nebula have been used to explain discrepancies
in abundances found from recombination lines
versus abundances found from collisionally-excited lines.
Using solely lines originating from the O$^{++}$ gas, we derive $t^2$ for the 
nebula
($t^2 = 0.009\pm0.004$) and the shock ($t^2 = 0.010\pm0.010$). \citet{est04} have 
published
a series of $t^2$ for a number of ions, including O$^{++}$ ($0.020\pm0.002$), as 
well as an average
from their series of ions ($0.022\pm0.002$).  The interesting result is that the 
shock maintains
a $t^2$ similar to the nebula (albeit with a large uncertainty)
despite being much thinner.
These observations, if corroborated with higher $S/N$ data, may draw
into question some of the theories that have been expounded surrounding an 
explanation for these
inferred $t^2$ fluctuations.
Grains appear to be present in the shock,
suggesting that the grains may still somehow be contributing to $t^2$.
The measurement of a non-zero $t^2$ in a 
matter-bounded shock would more likely
support the argument for an ``effective'' $t^2$ resulting from uncertainties in 
the atomic data and/or 
missing contributions to the line emission.
Higher $S/N$ O~{\sc ii} spectra of the shock will reduce the 
uncertainty of the inferred
$t^2$, and allow for more definitive conclusions to be made.
\acknowledgements
This work was supported by the Natural Sciences and Engineering
Research Council of Canada.  Line wavelengths were obtained from the
 Atomic Line List\footnote{Atomic Line List v2.04 is available at:
http://www.pa.uky.edu/\~{}peter/atomic/.} maintained by P.~A.~M.~van~Hoof. 
Calculations were performed with version 05.07 of Cloudy, last described by 
\citet{fer98}.
The authors wish to thank C.~R.~O'Dell for his clarification of a portion of
the BOM data, and referee M.~Peimbert for his detailed review of this paper.

\appendix
\section{Revisiting [Fe~{\sc iii}] energy levels  \label{velgrad}}
\begin{figure} 
\epsscale{0.8}
\centerline{\rotatebox{270}{\scalebox{0.9}[0.9]{\plotone{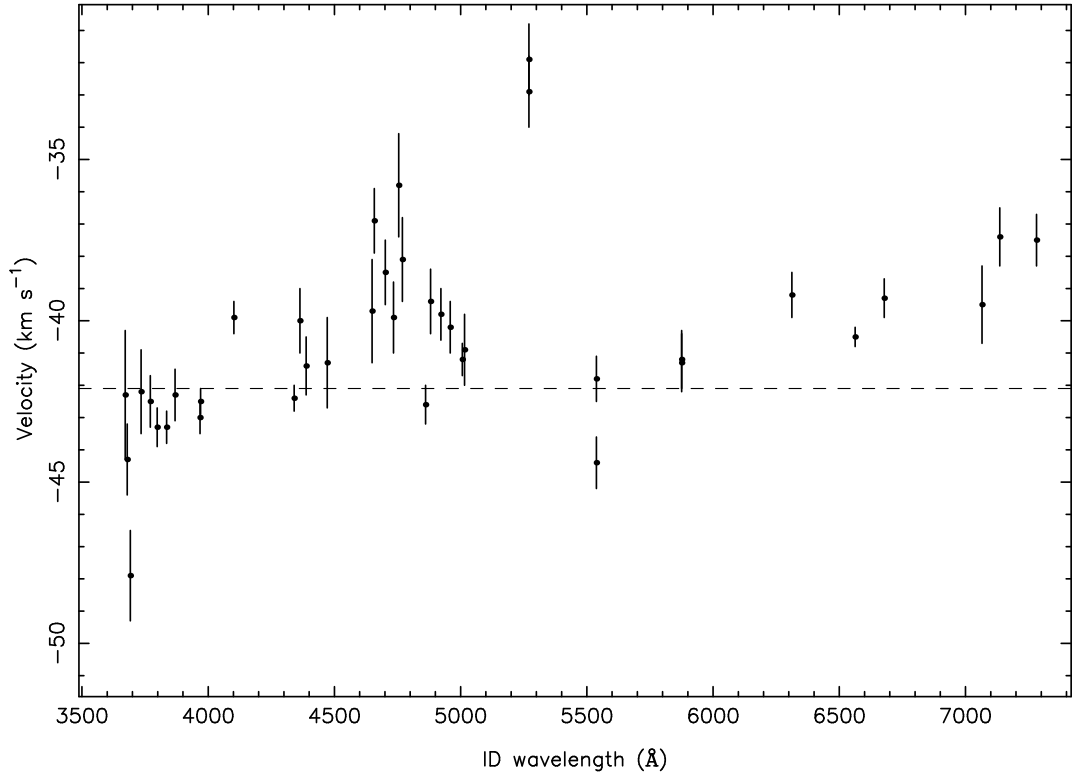}}}}
\caption{Measured velocity of shock component (with respect to nebular
H~{\sc i} recombination lines)
for high $S/N$ ($S/N > 5.2$) lines from unconstrained double Gaussian fit
with $1\sigma$ confidence interval determined
from the fit.  Constrained double Gaussian velocities
($-42.1$~km~s$^{-1}$) are not plotted individually, but are
represented by the horizontal dashed line.  The anomalous
[Fe~{\sc iii}]~5270.4 velocity is discussed in \S~\ref{analvel}.
\label{shockvel}}
\end{figure}

\begin{figure}
\epsscale{0.8}
\centerline{\rotatebox{270}{\scalebox{0.9}[0.9]{\plotone{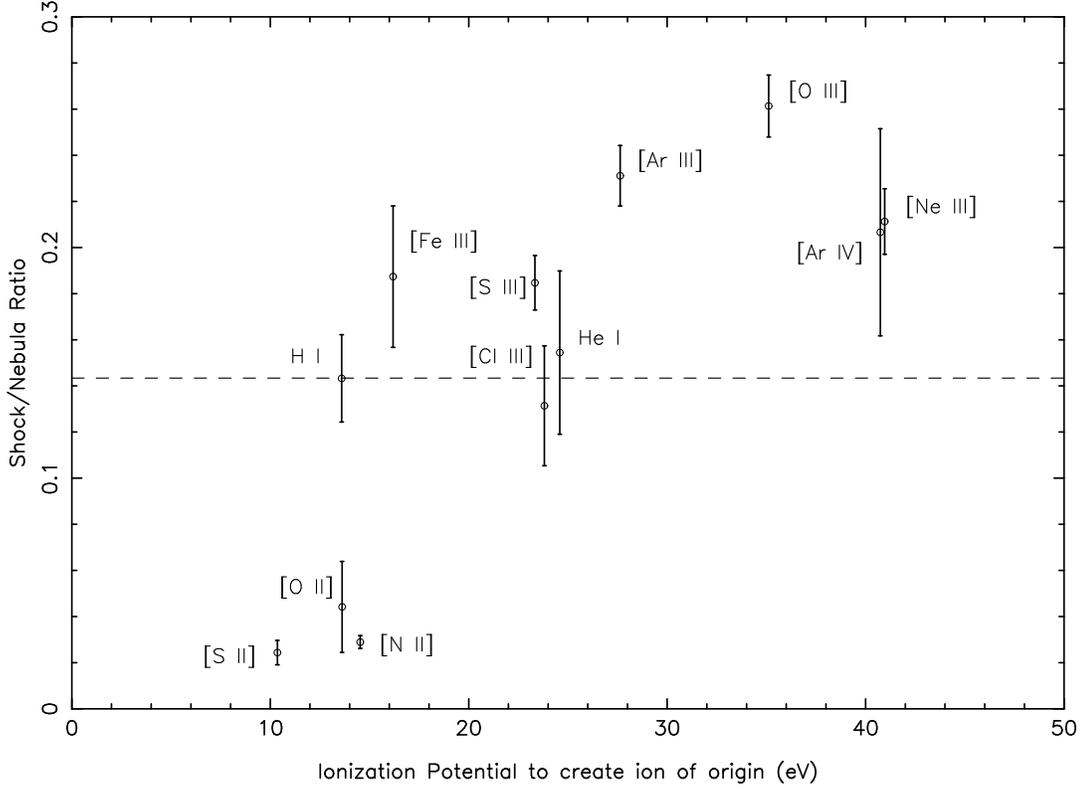}}}}
\caption{Weighted average of shock-to-nebula line ratios
(with $1\sigma$ error bars) from Column (9) of
Table~\ref{lines} as a function of the ionization potential needed to
create the associated originating ion.  The ionization potential serves to
differentiate ions with different ionization fractions while the
shock-to-nebula ratio is a measure of the optical thickness of the shock
to ionizing radiation.  If the
shock were optically thick, this
ratio would be close to one for all lines.
H~{\sc i} is presented as a standard for shock/nebula ionization comparison 
as its
originating ion (H$^{+}$) has an ionization fraction of roughly one
throughout
both the shock and the nebula.
The ratios of the medium-ionization species
([O~{\sc iii}], [Ar~{\sc iii}], [Ne~{\sc iii}]) all lie
above H~{\sc i} as their ionization fractions (averaged through the model) are higher
in the shock than in the nebula.
Conversely, the ratios of the low-ionization species
([O~{\sc ii}], [N~{\sc ii}], [S~{\sc ii}]) lie below H~{\sc i}.
\label{ratio}}
\end{figure}

\begin{figure} 
\epsscale{1.0}
\centerline{\rotatebox{270}{\scalebox{1.0}[1.0]{\plottwo{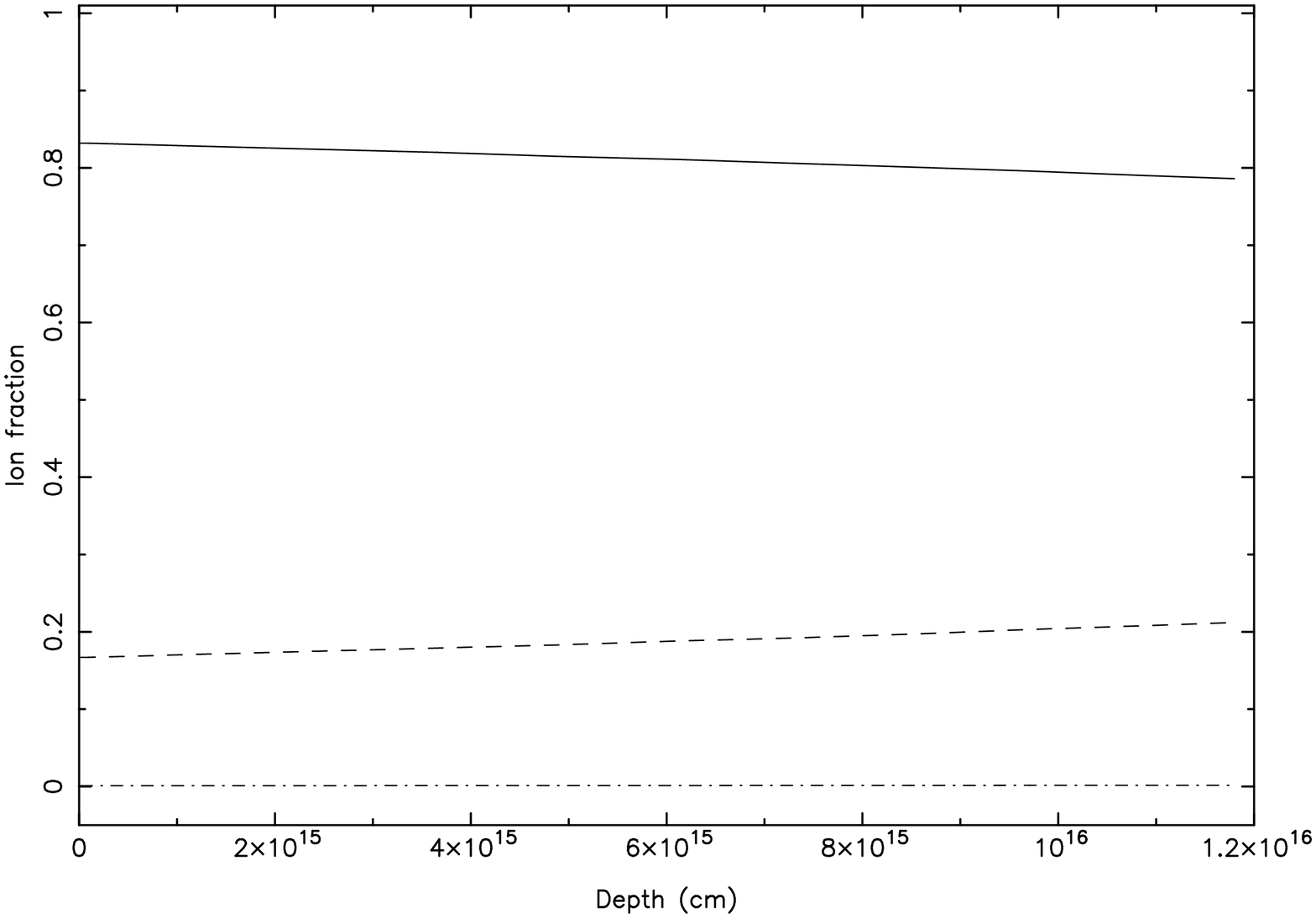}{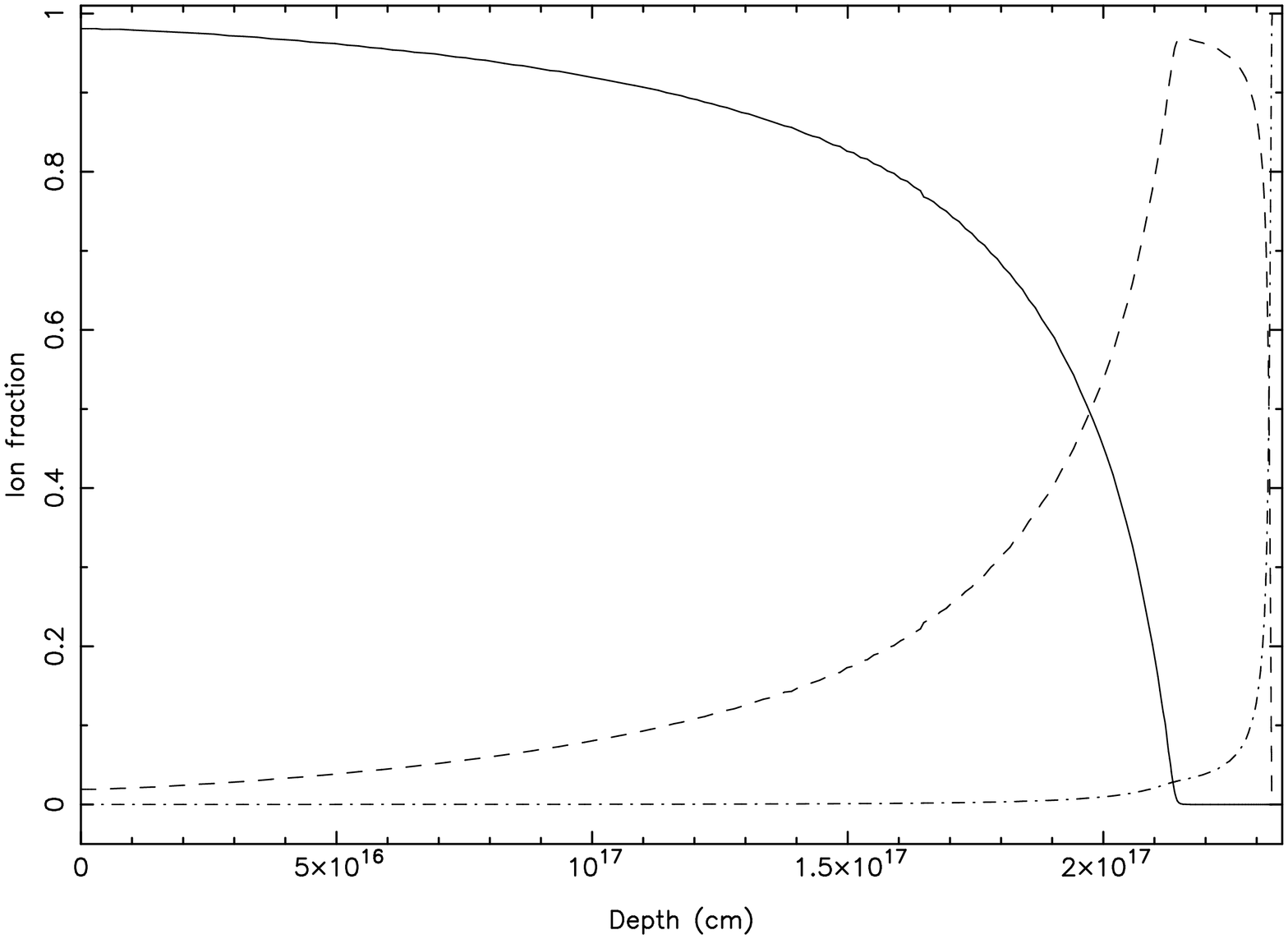}}}}
\caption{Ionization fraction of Fe as a function of depth from the
x2shock model (top) and from the nebular model (below) Fe$^{+++}$~-----;
Fe$^{++}$~-~-~-~- ; Fe$^{+}$~-$\cdot$-$\cdot$-$\cdot$- .  These structure
differences lead to differences between
shock-to-nebula ratios for low- and medium-ionization species as shown in
Fig~\ref{ratio}.
\label{ionizeiron}}
\end{figure}

As there is no velocity gradient in the shock,
all shock lines should have the same velocity relative to the H~{\sc i} lines
in the nebula.
However, a velocity discrepancy associated with the [Fe~{\sc
iii}]~5270 line appears in Figure~\ref{shockvel}.
The most obvious explanation for this is that the ID
wavelength for the [Fe~{\sc iii}]~5270 is wrong because of an error in
the adopted energy
of the upper $^3$P$^4_2$ energy level.  The other line originating from the
same upper level, [Fe~{\sc iii}]~5412 is much weaker and
therefore the velocity of the shock component cannot be measured as
reliably.  However,
a constrained Gaussian with a shock velocity that is consistent with that
observed in [Fe~{\sc iii}]~5270 does appear to fit the data well, albeit
with a $S/N\sim3$ for the shock component.
The $^3$P$^4_2$ energy is
quoted as $19404.8\pm0.5$~cm$^{-1}$ \citep{sug85}.  This uncertainty
translates to $\pm0.14$\AA, or $\pm7.9$~km~s$^{-1}$ for both [Fe~{\sc
iii}]~5270.40 and
[Fe~{\sc iii}]~5411.98.
The red and blue observations were used to constrain the energy of
this common upper level, 
fixing the energies of the lower levels at their
NIST values\footnote{http://physics.nist.gov/PhysRefData/ASD/index.html}.
For these lines to have a velocity consistent with that of the shock
(-42.1~km~s$^{-1}$),
$^3$P$^4_2$ must be
$19404.44\pm0.26$~cm$^{-1}$.

This is interesting in the context of the Fe$^{++}$ velocity
gradient presented in \citet{bal00}.  A velocity gradient of [Fe~{\sc iii}]  
initially observed as a function of wavelength was re-interpreted to
be a velocity gradient as a function of the lines' upper level
excitation potential above the ground state.  The
interpretation was presented
with scepticism as there was no evidence (or explanation)
for velocity gradients associated with any other single ion.
 The adjustment of the $^3$P$^4_2$ term
lowers the nebular velocity of [Fe~{\sc iii}]~5270 and [Fe~{\sc
iii}]~5412 to that
of the other [Fe~{\sc iii}] lines ($\sim4$~km~s$^{-1}$),
removing most of the evidence for a velocity gradient.  Note that this
nebular velocity
is consistent with what is expected from the relationship between velocity
and ionization potential \citep{bal00}.

The only remaining evidence for a sharp velocity gradient in the
Fe$^{++}$ zone is from the lines with $^3$G$_4$ (24940.9~cm$^{-1}$, \citep{sug85})
as their common upper
level ([Fe~{\sc iii}]~$4008$, [Fe~{\sc iii}]~$4080$).
The same uncertainty ($\pm0.5$~cm$^{-1}$) exists for this level,
translating to
uncertainties in the ID wavelengths of [Fe~{\sc iii}]~$4008.35\pm0.08$ and
[Fe~{\sc iii}]~$4079.70\pm0.08$, or equivalently to an uncertainty in the
velocity: $\pm6.1$~km~s$^{-1}$.  These two lines are too weak to be measured
in the shock, but for their line velocities
to be consistent with the nebula [Fe~{\sc iii}]
velocities, $^3$G$_4$ would have to be $24941.37\pm0.23$,
$\sim1\sigma$ above the accepted mean
\citep{sug85}.

In 
summary, if we require the concordance of [Fe~{\sc iii}] line
velocities in the shock and in the nebula, the $^3$P$^4_2$ energy would be
$19404.44\pm0.26$~cm$^{-1}$ producing lines with air wavelengths  
$5270.50\pm0.07$ and $5412.08\pm0.07$.
The $^3$G$_4$ energy would be $24941.37\pm0.23$
producing lines with wavelengths $4008.27\pm0.04$ and $4079.62\pm0.04$.

\clearpage
\begin{deluxetable}{rccrrrrrrl}
\tabletypesize{\tiny}
\tablewidth{0pt}
\tablecolumns{10}
\tablecaption{Summary of double Gaussian fits to observed echelle spectra. \label{lines}}
\tablehead{
\colhead{ID Wave\tablenotemark{a,b}} & 
\colhead{ID} & \colhead{?\tablenotemark{c}} & \colhead{Wavel\tablenotemark{d}} & 
\colhead{Velocity\tablenotemark{e}} & \colhead{FWHM\tablenotemark{e,f}} & \colhead{$I_{\rm corr}$\tablenotemark{g,h}} & \colhead{$S/N$\tablenotemark{i}} & 
\colhead{$I_{\rm sh}/I_{\rm neb}$\tablenotemark{j}} & 
\colhead{Notes\tablenotemark{k}} \\
\colhead{(1)} & \colhead{(2)} & \colhead{(3)} & 
\colhead{(4)} & \colhead{(5)} & \colhead{(6)} & 
\colhead{(7)} & \colhead{(8)} & \colhead{(9)} & 
\colhead{(10)} } 
\startdata
  3512.505 & \mbox{He~{\sc i}}     & ? &   3512.526 & $   1.8\pm1.7$ & $ 19.1\pm 4.4$ & $  0.0718\pm0.0167$ & 4.3 & $  0.6992\pm0.2724$ &  C               \\*
          &         & ? &   3512.023 & $ -41.1\pm1.7$ & $ 28.3\pm 0.0$ & $  0.0502\pm0.0157$ & 3.2 & & \\
   \nodata   &          & ? &   3530.556 &  \nodata      & $ 33.0\pm 9.9$ & $  0.1582\pm0.0546$ & 2.9 & \nodata & S               \\*
 & & & \nodata & \nodata & \nodata & \nodata & \nodata & & \\
  3554.389 & \mbox{He~{\sc i}}     &   &   3554.424 & $   3.0\pm0.8$ & $ 13.9\pm 1.8$ & $  0.1042\pm0.0129$ & 8.1 & \nodata &  S               \\*
 & & & \nodata & \nodata & \nodata & \nodata & \nodata & & \\
  3587.253 & \mbox{He~{\sc i}}     &   &   3587.281 & $   2.3\pm1.6$ & $ 23.9\pm 4.1$ & $  0.2617\pm0.0467$ & 5.6 & $  0.5319\pm0.1723$ &  C               \\*
          &         & ? &   3586.767 & $ -40.6\pm1.6$ & $ 28.3\pm 0.0$ & $  0.1392\pm0.0376$ & 3.7 & & \\
\enddata
\tablenotetext{a}{ID wavelength of line, in air.  Wavelengths are from the Atomic Line List v2.04 (http://www.pa.uky.edu/\~{}peter/atomic/, maintained by P.~A.~M.~van~Hoof), except \mbox{O~{\sc ii}} \citep{bla04}.}
\tablenotetext{b}{Nebular (neb) and shock (sh) components are included in separate, consecutive rows for each ID wavelength.
  If there is no shock component, the second row is left blank.
Lines with no identification are included, but Columns (1), (2) and (5) are left blank.}
\tablenotetext{c}{A low $S/N$ ($2.6 < S/N < 5.2$) is flagged with a `?'}
\tablenotetext{d}{Observed wavelength of line peak adjusted to rest frame of nebular H$^{+}$ as defined by median of first six unblended \mbox{H~{\sc i}} Balmer lines}
\tablenotetext{e}{km s$^{-1}$}
\tablenotetext{f}{A quoted FWHM of $28.3\pm0.0$ is the result of a constrained double Gaussian fit.}
\tablenotetext{g}{Reddening-corrected surface brightness ($I_{corr}$, $10^{-13}$ ergs cm$^{-2}$ s$^{-1}$ arcsec$^{-2}$) and its uncertainty calculated from the best-fit Gaussian area and
its $68.3\%$ confidence interval.}
\tablenotetext{h}{The red shock lines have been adjusted by
 a factor of 0.85 to correct for
 slit coverage differences between the red
 and blue spectra.  The nebular lines are not adjusted.}
\tablenotetext{i}{Signal-to-noise from surface brightness and $68.3\%$ confidence interval.}
\tablenotetext{j}{The shock-to-nebula ratio is entered within the `nebula' row and the `shock' row is left blank.}
\tablenotetext{k}{Constrained double Gaussian fit (C), constrained triple Gaussian fit (T), single Gaussian fit (S), red echelle spectrum (R), blue echelle spectrum (B), average of two orders (avg).}
\end{deluxetable}

\clearpage
\begin{deluxetable}{rcrrrrrcl}
\tabletypesize{\scriptsize}
\tablewidth{0pt}
\tablecolumns{9}
\tablecaption{Blue/red echelle spectra comparison. \label{redblue}}
\tablehead{
\colhead{ID Wave} & \colhead{ID} & \colhead{Wavel} &
\colhead{Velocity} & \colhead{FWHM} & \colhead{$I_{obs}$\tablenotemark{a}} &
\colhead{$S/N$} & \colhead{$B/R$\tablenotemark{b}} & \colhead{Notes} \\
\colhead{(1)} & \colhead{(2)} & \colhead{(3)} & 
\colhead{(4)} & \colhead{(5)} & \colhead{(6)} & 
\colhead{(7)} & \colhead{(8)} & \colhead{(9)} 
}
\startdata
  5270.400 & \mbox{[Fe~{\sc iii}]} &   5270.550 & $   8.5\pm0.1$ & $ 17.0\pm 0.3$ &     0.0248 & 57.5 & $     1.051\pm0.026$ & B avg    \\
 & &   5269.821 & $ -32.9\pm1.1$ & $ 25.6\pm 2.9$ &     0.0055 & 9.7 & $     0.846\pm0.115$ & \\
  5270.400 & \mbox{[Fe~{\sc iii}]} &   5270.555 & $   8.8\pm0.1$ & $ 16.1\pm 0.3$ &     0.0236 & 56.6 & \nodata & R               \\
 & &   5269.840 & $ -31.9\pm1.1$ & $ 27.6\pm 2.7$ &     0.0065 & 11.2 & \nodata & \\
  5517.720 & \mbox{[Cl~{\sc iii}]} &   5517.686 & $  -1.9\pm0.2$ & $ 15.6\pm 0.4$ &     0.0250 & 39.4 & $     1.232\pm0.048$ & B C             \\
 & &   5516.948 & $ -42.0\pm0.2$ & $ 28.3\pm 0.0$ &     0.0030 & 4.2 & $     0.789\pm0.237$ & \\
  5517.720 & \mbox{[Cl~{\sc iii}]} &   5517.697 & $  -1.3\pm0.2$ & $ 14.9\pm 0.5$ &     0.0203 & 33.8 & \nodata & R C             \\
 & &   5516.925 & $ -43.2\pm0.2$ & $ 28.3\pm 0.0$ &     0.0038 & 5.5 & \nodata & \\
\enddata
\tablenotetext{a}{$10^{-13}$ ergs cm$^{-2}$ s$^{-1}$ arcsec$^{-2}$, before reddening correction}
\tablenotetext{b}{The blue/red ratio ($B/R$) is entered in the `blue' rows for the nebula (1st row) and the shock (2nd row).  The `red' rows are left blank.}
\end{deluxetable}

\clearpage
\begin{deluxetable}{lccc}
\tabletypesize{\footnotesize}
\tablewidth{0pt}
\tablecolumns{4}
\tablecaption{Physical conditions within the nebula and the shock 
\label{temden_nebshock}}
\tablehead{
\colhead{Parameter} & \colhead{Line} & \colhead{Nebula} & \colhead{Shock}
}
\startdata
$N_e$ (cm$^{-3}$)
 & [S~{\sc ii}]\tablenotemark{a} & $5896^{+404}_{-366}$ & $13183^{+10000}_{-11183}$ \\
 & [O~{\sc ii}]\tablenotemark{b} & $1939^{+50}_{-50}$ & $2898^{+8429}_{-1997}$ \\
 & [O~{\sc ii}]\tablenotemark{c} & $3811^{+50}_{-50}$ & $7304^{+8429}_{-1997}$ \\
 & [Fe~{\sc iii}]\tablenotemark{d} & $4700^{+800}_{-800}$ & $7300^{+8000}_{-4100}$ \\
 & [Cl~{\sc iii}]\tablenotemark{e} & $12074^{+1300}_{-1118}$ & $21715^{+39170}_{-9641}$ \\
 & [Cl~{\sc iii}]\tablenotemark{f} & $7247^{+575}_{-519}$ & $10911^{+7515}_{-3665}$ \\
 & O~{\sc ii}\tablenotemark{g} & $6700^{+100}_{-100}$ & \nodata \\
$T_e$ (K) & [O~{\sc i}]\tablenotemark{h} & $8005^{+580}_{-408}$ & \nodata \\
 & [N~{\sc ii}]\tablenotemark{i} & $10672^{+53}_{-52}$ & $8784^{+1184}_{-729}$ \\
 & [O~{\sc iii}]\tablenotemark{j} & $8536^{+35}_{-33}$ & $8366^{+252}_{-214}$  \\
\enddata
\tablenotetext{a}{A: \citet{kee93} cs: \citet{ram96}}
\tablenotetext{b}{A: \citet{wie96} cs: \citet{mcl93}}
\tablenotetext{c}{A: \citet{zei82} cs: \citet{mcl93}}
\tablenotetext{d}{Weighted average from 11 [Fe~{\sc iii}] lines, following \citet{kee01}}
\tablenotetext{e}{A: \citet{kau86} cs: \citet{ram99}}
\tablenotetext{f}{A: \citet{men82} cs: \citet{ram99}}
\tablenotetext{g}{Following \citet{pei05}}
\tablenotetext{h}{A: \citet{wie96} cs: \citet{bha95}}
\tablenotetext{i}{A: \citet{wie96} cs: \citet{len94}}
\tablenotetext{j}{A: \citet{wie96} cs: \citet{len94}}
\end{deluxetable}

\clearpage
\begin{deluxetable}{lrrrrrrr}
\tablewidth{0pt}
\tabletypesize{\scriptsize}
\tablecolumns{8}
\tablecaption{O$^{++}$/H$^+$ ratios from permitted lines. \label{tempfluct}}
\tablehead{
\colhead{} & \colhead{} &
\multicolumn{3}{c}{Nebula} & \multicolumn{3}{c}{Shock} \\
\cline{3-5}  \cline{6-8} \\
\colhead{Transition}  & \colhead{Lines} & 
\colhead{$I_{sum}$\tablenotemark{a}} &
\multicolumn{2}{c}{O$^{++}$/H$^{+}$ ($\times10^{-5}$)} &
\colhead{$I_{sum}$\tablenotemark{a}} &
\multicolumn{2}{c}{O$^{++}$/H$^{+}$ ($\times10^{-5}$)} \\
\colhead{} & \colhead{(\AA)} & \colhead{($\times10^{-2}$)} &
\colhead{A\tablenotemark{b}} & \colhead{B\tablenotemark{c}} & 
\colhead{($\times10^{-2}$)} &
\colhead{A\tablenotemark{b}} & \colhead{B\tablenotemark{c}}
}
\startdata
 3s  $^4$P-3p $^4$D$^0$ & 4642+76 & $0.4285\pm0.0891$ & $27\pm6$ & $26\pm6$ &
				    $0.7381\pm0.1961$ & $47\pm12$ & $45\pm12$ \\
                        & 4649    & $0.4355\pm0.0213$ & $28\pm1$ & $27\pm1$ &
				    $1.0491\pm0.1525$ & $67\pm10$ & $65\pm10$ \\
\enddata
\tablenotetext{a}{Determined following \citet{pei05} (their equations 3 
and 4) using $N_e$(FL)$^{neb}\sim6000$, $N_e$(FL)$^{shock}\sim10000$}
\tablenotetext{b}{Case A O~{\sc ii} recombination coefficients \citep{sto94}}
\tablenotetext{c}{Case B O~{\sc ii} recombination coefficients \citep{sto94}}
\end{deluxetable}

\clearpage
\begin{deluxetable}{llrrr}
\tablewidth{0pt}
\tablecolumns{5}
\tablecaption{O/H from collisionally excited lines and 
recombination lines with a derived $t^2$ parameter. \label{RLCEL}}
\tablehead{
\colhead{} & \colhead{} & \multicolumn{2}{c}{12+log(X$^{m}$/H$^{+}$)} & 
\colhead{} \\
\cline{3-4} \\
 & \colhead{Feature} & \colhead{CELs\tablenotemark{a}} & 
\colhead{RLs\tablenotemark{b,c}} & 
\colhead{$t^{2}$}
}
\startdata
O$^{++}$ & Nebula & $8.37\pm0.01$ & $8.43\pm0.03$ & $0.009\pm0.004$ \\
 & Shock & $8.69\pm0.05$ & $8.76\pm0.08$ & $0.010\pm0.010$ \\
O$^{+}$ & Nebula & $7.78\pm0.03$ & \nodata & \nodata \\
 & Shock & $7.72\pm0.18$ & \nodata & \nodata \\
O$^{0}$ & Nebula & $6.19\pm0.03$ & \nodata & \nodata \\
 & Shock & \nodata & \nodata & \nodata \\
O$_{tot}$ & Nebula  & $8.47\pm0.01$ & $8.52\pm0.03$\tablenotemark{d} & \nodata \\
 & Shock & $8.73\pm0.05$ & $8.80\pm0.08$\tablenotemark{d} & \nodata
\enddata
\tablenotetext{a}{collisionally excited lines}
\tablenotetext{b}{recombination lines}
\tablenotetext{c}{using case B effective recombination coefficients \citep{sto94}}
\tablenotetext{d}{O$^{++}$ recombination; O$^{+}$, O$^{0}$ collisionally excited lines}
\end{deluxetable}

\clearpage
\begin{deluxetable}{lrrr}
\tabletypesize{\small}
\tablewidth{0pt}
\tablecolumns{4}
\tablecaption{Constraints on Model Parameters \label{constraint}}
\tablehead{
\colhead{Quantity} & 
\multicolumn{3}{c}{HH Shock} \\
\cline{2-4} \\
\colhead{} & \colhead{Obs} & 
\colhead{Mod A\tablenotemark{a}} & \colhead{Mod B\tablenotemark{b}} 
}
\startdata
 F(He~{\sc i}~$\lambda6678$)\tablenotemark{c} & $    0.31\pm0.01$ &     0.34 &     0.33 \\
 $\lambda5007/\lambda6678$\tablenotemark{d} & $  144.13\pm8.68$ &   136.45 &   149.67 \\
 $\lambda5007/\lambda3726$\tablenotemark{d} & $   35.35\pm6.96$ &    40.11 &    33.62 \\
 $\lambda6312/\lambda6725$\tablenotemark{d} & $   1.816\pm0.270$ &    0.851 &    1.421 \\
 $(\lambda4959+\lambda5007)/\lambda4363$\tablenotemark{e} & $  361.02\pm39.15$ &   381.57 &   380.17 \\
 $(\lambda6548+\lambda6583)/\lambda5755$\tablenotemark{e} & $   69.88\pm17.84$ &    78.84 &    70.48 \\
 $\lambda6731/\lambda6716$\tablenotemark{f} & $    2.01\pm0.59$ &     2.14 &     2.14 \\
 $\lambda3726/\lambda3729$\tablenotemark{f} & $    2.29\pm0.69$ &     2.57 &     2.57 \\
 $\lambda5518/\lambda5538$\tablenotemark{f} & $    0.50\pm0.10$ &     0.45 &     0.45 \\
\enddata
\tablenotetext{a}{Mihalas stellar atmosphere, CLOUDY H~{\sc ii} abundances}
\tablenotetext{b}{Kurucz stellar atmosphere, \citet{est04} abundances}
\tablenotetext{c}{$10^{-13}$ ergs cm$^{-2}$ s$^{-1}$ arcsec$^{-2}$}
\tablenotetext{d}{Ionization indicator}
\tablenotetext{e}{Temperature indicator}
\tablenotetext{f}{Density indicator}
\end{deluxetable}

\clearpage
\begin{deluxetable}{lrr}
\tablewidth{0pt}
\tablecolumns{3}
\tablecaption{CLOUDY input parameters\label{cloudy}}
\tablehead{   
\colhead{Quantity} &
\colhead{Shock Mod A\tablenotemark{a}} & \colhead{Shock Mod
B\tablenotemark{b}}
}
\startdata
$T_{\mathrm{star}}$ (K) & 35200 & 41200 \\
log$\phi$(H) & 12.700 & 12.800 \\
radius (pc) & undef & undef \\
log$n_{\mathrm H}\mathrm{(inner)}$ & 4.20 & 4.20\\
Constant & $\rho$ & $\rho$ \\
Grains & yes & yes \\
thickness (pc) &  0.00396 &  0.00391 \\
\enddata
\tablenotetext{a}{Mihalas stellar atmosphere, CLOUDY H~{\sc ii} abundances}
\tablenotetext{b}{Kurucz stellar atmosphere, \citet{est04} abundances}
\end{deluxetable}

\clearpage
\begin{deluxetable}{lrr}
\tabletypesize{\small}
\tablewidth{0pt}
\tablecolumns{3}
\tablecaption{Model abundances relative to H (12+log($X$/H))\label{abund_table}}
\tablehead{
\colhead{Element} &
\colhead{Shock Mod A\tablenotemark{a}} & \colhead{Shock Mod B\tablenotemark{b}} \\
\colhead{(1)} & \colhead{(2)} & \colhead{(3)}
}
\startdata
He &      10.98 &      10.98 \\
C  &       8.48 &       8.42 \\
N  &       7.85 &       7.73 \\
O  &       8.60 &       8.65 \\
Ne &       7.78 &       8.05 \\
S  &       7.00 &       7.22 \\
Ar &       6.48 &       6.62 \\
Cl &       5.00 &       5.46 \\
\enddata
\tablenotetext{a}{CLOUDY H~{\sc ii} region abundances from \citet{bal91,rub91,ost92}}
\tablenotetext{b}{\citet{est04} abundances}
\end{deluxetable}

\clearpage
\begin{deluxetable}{crrrr}
\tabletypesize{\footnotesize}
\tablewidth{0pt}
\tablecolumns{5}
\tablecaption{Comparison of model predictions and observations. \label{modeltab}}
\tablehead{
\colhead{ID} & \colhead{ID} & \multicolumn{3}{c}{HH Shock\tablenotemark{a}} \\
\cline{3-5} \\
\colhead{Species} & \colhead{Wavelength} & \colhead{Obs} & \colhead{Mod A\tablenotemark{b}} & \colhead{Mod B\tablenotemark{c}}
}
\startdata
\mbox{He~{\sc i}} & 3512.5 & $   0.164\pm0.051$ &    0.039 &    0.039 \\
\mbox{He~{\sc i}} & 3587.3 & $   0.455\pm0.123$ &    0.096 &    0.096 \\
\mbox{He~{\sc i}} & 3613.6 & $   0.154\pm0.041$ &    0.130 &    0.111 \\
\mbox{He~{\sc i}} & 3634.2 & $   0.138\pm0.051$ &    0.137 &    0.138 \\
\mbox{H~{\sc i}} & 3667.7 & $   0.169\pm0.030$ &    0.251 &    0.172 \\
\mbox{H~{\sc i}} & 3671.5 & $   0.190\pm0.035$ &    0.279 &    0.192 \\
\mbox{H~{\sc i}} & 3673.8 & $   0.124\pm0.028$ &    0.313 &    0.215 \\
\mbox{H~{\sc i}} & 3676.4 & $   0.203\pm0.032$ &    0.352 &    0.243 \\
\enddata
\tablenotetext{a}{All fluxes are quoted relative to He~{\sc i}~6678.  In the cases where lines were observed in both the red and the blue spectrum, the blue observations are listed first, and the model predictions are not repeated.} 
\tablenotetext{b}{Mihalas stellar atmosphere, CLOUDY H~{\sc ii} abundances} 
 \tablenotetext{c}{Kurucz stellar atmosphere, \citet{est04} abundances} 
 \tablenotetext{d}{Measurement is a blend} 
  \tablenotetext{e}{Blue echelle spectrum}
 \tablenotetext{f}{Red echelle spectrum}
 \end{deluxetable}

\clearpage
\begin{deluxetable}{lrrrrrcrr}
\tabletypesize{\footnotesize}
\tablewidth{0pt}
\tablecolumns{9}
\tablecaption{Iron abundance and depletion in the Orion nebula and HH shock \label{depletion_table}}
\tablehead{
\colhead{Element} & \multicolumn{5}{c}{Nebula} & & \multicolumn{2}{c}{Shock}  \\
\cline{2-6} \cline{8-9} \\
 & \colhead{B91\tablenotemark{a}} & \colhead{R97\tablenotemark{b}} & \colhead{E98\tablenotemark{c}} & \colhead{E04\tablenotemark{d}} &
\colhead{CLOUDY\tablenotemark{e}} & & \colhead{Mod A} & \colhead{Mod B}
}
\startdata
Gas (Fe/H) & 0.042e-4 & 0.0016e-4 & 0.026e-4 & 0.017e-4 & 0.03e-4 & & 0.034e-4 & 0.0523e-4 \\
Gas (12+log(Fe/H)) & 6.62 & 5.20 & 6.41 & 6.23 & 6.48 & & 6.53 & 6.72 \\
Depletion\tablenotemark{f} (dex) & -0.88 & -2.30 & -1.08 & -1.27 & -1.02 & & -0.97 & -0.78 \\
\enddata
\tablenotetext{a}{\citet{bal91}}
\tablenotetext{b}{\citet{rub97}, noted to be an anomalously low Fe abundance}
\tablenotetext{c}{\citet{est98}}
\tablenotetext{d}{\citet{est04}}
\tablenotetext{e}{CLOUDY H~{\sc ii} region abundances from \citet{bal91,rub91,ost92}}
\tablenotetext{f}{Relative to solar (7.50, \citet{gre99})}
\end{deluxetable}

\clearpage

\clearpage

\clearpage

\clearpage

\clearpage

\clearpage

\clearpage

\clearpage

\end{document}